\documentclass[10pt,iop]{emulateapj}

\usepackage{graphicx}
\usepackage{float}
\usepackage{amsmath}
\usepackage{epsfig}
\usepackage{color}

\newcommand{\ve}[1]{\mathbf{#1}}

\newcommand{\lm}{{\ell m}}

\newcommand{\A}{\ve{A}}
\newcommand{\D}{\ve{D}}
\newcommand{\F}{\ve{F}}
\renewcommand{\S}{\ve{S}}
\newcommand{\lmax}{{\ell_\text{max}}}
\newcommand{\lmaxh}{{\ell^h_\text{max}}}
\newcommand{\N}{\ve{N}}
\newcommand{\M}{\ve{M}}

\newcommand{\B}{\ve{B}}
\newcommand{\Y}{\ve{Y}}
\newcommand{\Yobs}{\ve{Y}_\text{obs}}

\renewcommand{\d}{\ve{d}}
\newcommand{\x}{\ve{x}}
\newcommand{\e}{\ve{e}}
\renewcommand{\r}{\ve{r}}
\renewcommand{\b}{\ve{b}}
\newcommand{\s}{\ve{s}}
\newcommand{\n}{\ve{n}}
\newcommand{\W}{\ve{W}}
\newcommand{\I}{\ve{I}}

\newcommand{\Nsideh}{N^h_{\text{side}}}

\newcommand{\Nside}{N_\text{side}}
\newcommand{\Npix}{N_\text{pix}}

\begin{document}

\title{A multi-level solver for Gaussian constrained CMB realizations}

\author{D. S. Seljebotn\altaffilmark{1},
K.-A. Mardal\altaffilmark{2,3},
J. B. Jewell\altaffilmark{4},
H. K. Eriksen\altaffilmark{1},
and P. Bull\altaffilmark{1}}

\email{d.s.seljebotn@astro.uio.no}

\altaffiltext{1}{Institute of Theoretical Astrophysics, University of
  Oslo, P.O.\ Box 1029 Blindern, N-0315 Oslo, Norway}

\altaffiltext{2}{Department of Informatics, University of Oslo, P.O.\ Box 1080 Blindern, N-0316 Oslo, Norway}
\altaffiltext{3}{Centre for Biomedical Computing, Simula Research Laboratory, P.O.\ Box 134, N-1325 Lysaker, Norway}

\altaffiltext{4}{Jet Propulsion Laboratory, California Institute of Technology, Pasadena, CA 91109, USA}

\begin{abstract}
  We present a multi-level solver for drawing constrained Gaussian
  realizations or finding the maximum likelihood estimate of the CMB
  sky, given noisy sky maps with partial sky coverage. The method
  converges substantially faster than existing Conjugate Gradient (CG)
  methods for the same problem. For instance, for the 143 GHz Planck
  frequency channel, only 3 multi-level W-cycles result in an absolute
  error smaller than 1$\,\mu\textrm{K}$ in any pixel. Using 16 CPU
  cores, this translates to a computational expense of 6 minutes wall
  time per realization, plus 8 minutes wall time for a power 
  spectrum-dependent precomputation. Each additional W-cycle reduces 
  the error by more than an order of magnitude, at an additional computational
  cost of 2 minutes.  For comparison, we have never been able to
  achieve similar absolute convergence with conventional CG methods
  for this high signal-to-noise data set, even after thousands of CG
  iterations and employing expensive preconditioners. The solver is
  part of the Commander 2 code, which is available with an open source
  license at \verb@http://commander.bitbucket.org/@.

\end{abstract}
\shorttitle{Multi-level solver for constrained CMB realizations}

\section{Introduction}

Apart from a substantial kinematical dipole, the cosmic microwave background 
(CMB) radiation is observed to be isotropic to around one part in $10^4$. Below 
this level, there are random fluctuations over a wide range of angular scales. 
The prevailing `concordance' cosmological model explains these anisotropies as 
the imprints of Gaussian-distributed, statistically-isotropic perturbations of 
spacetime that were generated during an inflationary epoch in the early 
Universe. Correlations between the fluctuations provide a wealth of information 
about inflation and the subsequent growth of structure, and so being able to 
accurately measure and characterize them is of paramount importance to modern 
cosmology.

As detector technology has improved, it has become possible to probe smaller 
and smaller angular scales with ever-increasing noise sensitivities. The 
resulting improvement in resolution and signal-to-noise ratio presents a 
formidable computational challenge, as one must now reliably reconstruct the 
CMB sky to high accuracy over tens of millions of pixels, while simultaneously 
taking into account complexities of the data such as inhomogeneous noise, 
foreground contamination, and regions of missing/masked data.

Consider an observed map of the CMB,
for instance similar to those provided by the WMAP
\citep{bennett:2012} and Planck \citep{planck:2013a} experiments. The
ideal CMB map would consist of an error-free value at every single
position on the sky. In reality this is of course not
possible, because of instrumental imperfections 
(such as noise and beam smoothing) and
strong foreground contamination from astrophysical sources; there will
always be uncertainties in a real CMB map. Therefore, rather than
aiming to extract ``a single true CMB sky map'', a more realistic solution is
to compute an ensemble of many possible CMB skies, each of which is
both noise-free, full-sky, and \emph{statistically consistent} 
with the observed data. 
This idea has already been implemented for CMB analysis purposes 
in terms of a Gibbs sampling framework, as described by
\citet{jewell:2004,wandelt:2004,eriksen:2004, eriksen:2008a}.

An underlying assumption in this line of work is that both the CMB sky
and instrumental noise are random Gaussian fields with covariance
matrices $\mathbf{S}$ and $\mathbf{N}$, respectively. In most
applications -- following the basic inflationary prediction -- one 
additionally assumes that the CMB field is
isotropic, so that the CMB covariance matrix can be specified in
terms of a simple angular power spectrum, $C_{\ell}$. Of course, this
power spectrum is not known {\em a priori}, but must instead be
estimated from the data, and indeed, this is usually the main goal for
most CMB experiments.

The Gibbs sampling framework provides a well-structured mathematical
solution to this power spectrum estimation problem, by establishing
the full joint Bayesian posterior distribution of the CMB sky and CMB
power spectrum. This is found by iteratively sampling from the
(more tractable) conditional distributions according to a simple 
algorithm: 1) Make an arbitrary initial `guess' for the CMB power 
spectrum; 2) draw a CMB sky map compatible
with the data and the assumed power spectrum; 3) draw a power spectrum
compatible with the sky sample that was just drawn; and 4) iterate. 
The resulting set of sky and power spectrum samples will (after 
some burn-in period) converge to the true joint posterior distribution.

Although simple to write down, this algorithm is also computationally
rather expensive due to step (2), which
essentially amounts to solving a large linear system with one or more
random terms on the right-hand side, corresponding to different
realizations. We will refer to this system as the \emph{constrained
  realization} (CR) system. The same linear system can also be solved
for the maximum likelihood CMB sky map estimate, which is sometimes
referred to as the {\em Wiener-filtered map}.  Since the degrees of
freedom of the CR system scale with the number of pixels, brute force
solutions are out of bounds except for very low-resolution data
sets. However, it is computationally feasible to multiply an arbitrary
vector with the system matrix by repeatedly changing basis functions
(i.e. spherical harmonic transforms), so that the system can be
solved using iterative linear equation solvers. The main problem is to
optimize the convergence rate of these solvers to produce a solution
in a timely manner. 

Commander \citep{eriksen:2004}, the CMB Gibbs sampler mentioned above,
solves the CR system through the Conjugate Gradient (CG) method, using
a combination of a block preconditioner on large angular scales and a
diagonal preconditioner on small angular scales. While this approach
was successful for analyzing WMAP observations
\citep{odwyer:2004,eriksen:2007a,eriksen:2007b,eriksen:2008b}, the
higher signal-to-noise level of data from more recent 
experiments like Planck effectively halts convergence of the solver.
Indeed, as we will see in Section
\ref{sec:conditioning}, the number of CG iterations intrinsically
scales with the signal-to-noise ratio of a given data set, 
limiting the utility of CG for data sets such as these. To 
produce the low-$\ell$ power spectrum likelihood for the Planck 
mission, for example, the data had to be downgraded to low angular 
resolution and a substantial amount of regularization noise added 
\citep{planck:2013b}. Even then, several thousands of CG iterations
were required for convergence. To go to full angular resolution with
this scheme is simply not computationally feasible.

A somewhat better approach was described by \cite{smith:2007}, who
applied the CG method recursively, such that a CG solution on a coarse
grid was used as the preconditioner for CG on a finer grid.  We are
not aware of any head-to-head comparisons of this method versus the
one described by \cite{eriksen:2004}, but our understanding is that,
although it is faster, it still scales with the signal-to-noise ratio
of the data set, and therefore does not inherently fix the fundamental
convergence problems for high-sensitivity, high-resolution 
analysis.

More recently, \cite{elsner:2012,elsner:2013} introduced a stationary
iterative method for solving the CR equation. They did not quote the 
usual statistics for convergence, such as total reduction in residual 
and error, however. Not knowing the accuracy of their solution, we are
unable to compare the efficiency of their method directly to ours. While 
they do quote the change in the $\chi^2$ statistic of the posterior 
probability density {\em between successive iterations}, iterative
methods (and stationary methods in particular) are vulnerable to
breaking down in terms of convergence rate well before reaching true
convergence.  Also, the $\chi^2$ explicitly ignores large scales under
the mask. While there certainly are applications where this is acceptable,
CMB Gibbs sampling is not one of them, since it explicitly
iterates between considering the CMB signal a sample from the
posterior, which mostly ignores the masked area, and a sample from the
prior, which gives equal weight to the masked area.

In this paper we present a new solver for the CR system that is
radically different from the CG approach, and instead builds on the
multi-level (or multi-grid) framework. These algorithms are best known
in the astrophysics community as solvers for elliptical partial
differential equations (PDEs), although they are in fact more generally 
applicable to solving many types of linear systems \citep{brandt:2001}. 
We apply multi-level theory to the CR equation (although the algorithm is not
entirely traditional), and show that the resulting algorithm converges
to the exact solution with only a handful of iterations even for the
most sensitive Planck channel. Most importantly, and contrary 
to the CG solver, the convergence rate is nearly independent of the
signal-to-noise ratio of the data set. 

Multi-level methods have been explored before in the CMB community for
the purposes of map-making. \cite{dore:2001} described a standard
multi-grid method for map-making, although it was eventually unable
to compete with standard CG and approximate map-makers.
\cite{grigori:2012} also presented a promising two-level CG preconditioner
for map-making based on the domain-decomposition method in
\cite{have:2013}. The map-making equation is different from
CR equation, however, in that one does not solve for the CMB signal under a
mask. As we will see in Section \ref{sec:conditioning}, it is this feature
in particular that makes convergence difficult to achieve on the CR
system.

\setcounter{footnote}{0}

\section{Exploring the CR linear system}

\subsection{Matrix notation for spherical harmonic transforms}
\label{sec:sht}

The details of changing between pixel domain and spherical harmonic
domain are usually glossed over in the literature. Since we
will be solving a large linear system that couples
signals on all scales --- from individual pixels to the full sky --- it is
of the utmost importance to be precise about how these conversions are
performed. If implemented incorrectly, even small pixel-scale errors
can lead to overall divergence of the entire method.

There is no perfect grid on the sphere, and in choosing a particular
one, a number of trade-offs must be considered. In our current
implementation we adopt both the
HEALPix\footnote{http://healpix.sourceforge.net} pixelization
\citep{gorski:2005} and the Gauss-Legendre spherical grid \citep[and
  references therein]{libpsht}. The HEALPix software package contains
routines that are useful for our pixel domain computations, while the
latter is required for accurate evaluation of Equation
\eqref{eq:sh-analysis} below.

Given such a grid on the sphere (by which we mean a set
of positions $\hat{n}_i$ on the sky), we can use \emph{spherical
  harmonic synthesis} to transform a field expressed in
spherical harmonic basis, with coefficients $s_\lm$, to a field
sampled on the sphere,
\begin{equation}
  \label{eq:sh-synthesis}
  \hat{s}(\hat{n}_i) = \sum_{\ell = 0}^\lmax \sum_{m=-\ell}^\ell s_\lm
  Y_\lm(\hat{n}_i).
\end{equation}
We will write this operation in matrix form as $\widehat{\s} = \Y \s$,
where $\Y$ encodes the value of the spherical harmonics evaluated at
each $\hat{n}_i$ of the chosen grid. Note that $\Y$ is not a square
matrix, as spherical grids need to over-sample the signal to
faithfully represent it up to some bandlimit $\lmax$. In typical
applications there are between $30\%$ and $100\%$ more pixels along
the rows of $\Y$ than there are spherical harmonic coefficients along
the columns. For the purposes of our method, it will turn out that we 
need to under-pixelize the signal instead, so there will be more columns
than rows in $\Y$.

The opposite action of converting from pixel basis to harmonic basis
is \emph{spherical harmonic analysis}, which generally takes
the quadrature form
\begin{equation}
  \label{eq:sh-analysis}
s_{\ell m} = \int_{4\pi} Y_\lm^*(\hat{n}) \hat{s}(\hat{n}) d\Omega
\approx \sum_{i=1}^{\Npix} Y_\lm^*(\hat{n}_i) w_i \hat{s}(\hat{n}_i),
\end{equation}
where $w_i$ combines quadrature weights and pixel area. Similar to the
synthesis case, this operation can be written in matrix form as
$\s=\Y^T\W~\widehat{\s}$, where $W_{ij} = w_i \delta_{ij}$.
A crucial feature of our method is the ability to (for the
most part) avoid spherical harmonic analysis, however. Instead, we will rely on
{\em adjoint spherical harmonic synthesis}, $\Y^T$, which simply
appears algebraically as the transpose of $\Y$.

Note that, unlike in the case of the more famous discrete Fourier
transform, $\Y$ is not a square orthogonal matrix, and synthesis and
analysis differ by more than transposition and a scale factor. One may
in some situations have that $\Y^T \W \Y = \ve{I}$, but this depends
on both $\lmax$, $\Npix$ and the spherical grid.

The action of applying $\Y$, $\Y^T$, $\Y^T\W$ or $\W \Y$ to a vector
is in general referred to as a {\em spherical harmonic transform}
(SHT). Carefully-optimized libraries are available that perform SHTs
in $O(\lmax \Npix)$ time; we use the \verb@libsharp@ library
\citep{reinecke:2013}.

\subsection{Data model}
\label{sec:data-model}

We now define our data model, and assume from the
beginning that the CMB is Gaussian and isotropic
\citep[e.g.][]{planck:2013c}. Following the notation of
\citet{eriksen:2004}, it is convenient to define the CMB signal to be
a vector $\s$ of spherical harmonic coefficients, in which case the
associated covariance matrix $\S$ is given by
\[
S_{\ell m,\ell' m'} = \delta_{\ell \ell'} \delta_{mm'} C_\ell,
\]
where $C_\ell$ is the CMB power spectrum.

Using the notation of the previous section, the model for the observed
sky map pixel vector, $\d$, is
\begin{equation}
  \label{eq:Y-data-model}
  \d = \Yobs \B \s + \n,
\end{equation}
where $\B$ denotes beam-smoothing and the pixel window function, $\n$
is Gaussian instrumental noise, and the subscript of $\Yobs$ indicates
projection to the pixelization of the map $\d$.

We assume a symmetric instrumental beam, so that the beam matrix $\B$
is a diagonal matrix given by $B_{\ell m,\ell' m'} = b_\ell p_\ell
\delta_{\ell \ell'} \delta_{mm'}$, where $b_\ell$ is the instrumental
beam and $p_\ell$ the pixel window function of the observed grid. We
also assume white instrumental noise, such that the noise covariance 
matrix, $\N$, is diagonal. We discuss the likely impact of asymmetric 
beams and correlated noise in Section \ref{sec:discussion}.

Discretization of the model is done simply by picking some $\lmax$ for
the $\s$ vector. The noise vector $\n$ is related to the map-making
process, averaging the noise of time-ordered data (TOD) that fall
within the same pixel, and so is inherently discrete rather than being a
discretization of any underlying field. As already mentioned above,
no spherical harmonic analysis of $\d$ (and therefore $\n$) is required 
when solving the CR system; rather, one solves for the projected
$\s$, and so the noise treatment is always perfectly consistent
with the assumed model.

\subsection{The CR linear system}
\label{sec:cr-linear-system}

Given the data model above, we are interested in exploring the
Bayesian posterior distribution $p(\s | \d, C_\ell)$, the CMB signal
given the data and CMB power spectrum.  Let us first define
\begin{equation}
  \label{eq:system-explicit}
  \A \equiv \S^{-1} + \B \Yobs^T \N^{-1} \Yobs \B,
\end{equation}
where in what follows we will refer to the first term as the {\em
prior term}, and the second as the {\em inverse-noise term}. It 
can be shown that if we now solve the {\em CR system}
\begin{equation}
  \label{eq:cr-system}
  \A  \x = \B \Yobs^T \N^{-1} \d,
\end{equation}
the solution $\x$ will be the maximum likelihood estimate of $\s$. 
Alternatively, if particular random fluctuation terms are added 
to the right-hand side of Eq. \eqref{eq:cr-system}, the solution $\x$
will instead be samples from the posterior
\citep{jewell:2004,wandelt:2004}. Since $b_\ell \to 0$ as $\ell$
increases, the diagonal prior term will at some point dominate the
dense inverse-noise term, so that truncation at sufficiently high
$\lmax$ does not affect the solution of the system.

\begin{figure}
  \begin{center}
    \includegraphics[width=\linewidth]{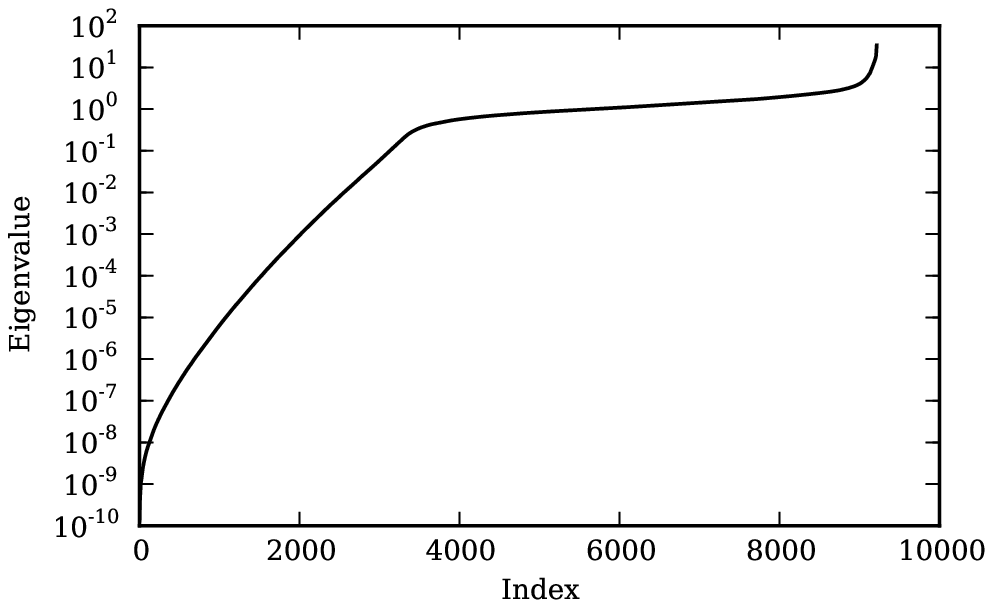}
    \includegraphics[width=\linewidth]{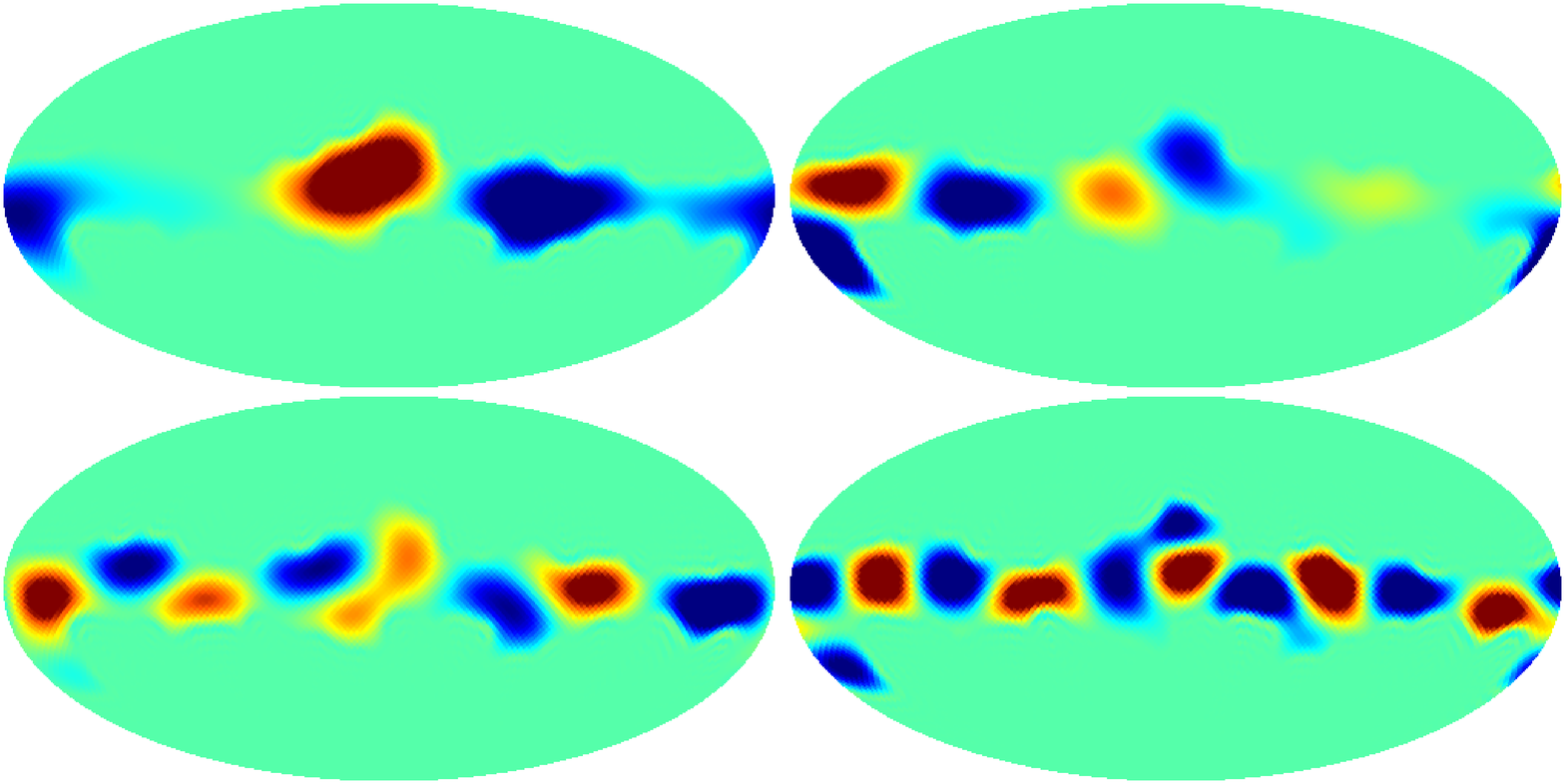}
    \includegraphics[width=0.5\linewidth]{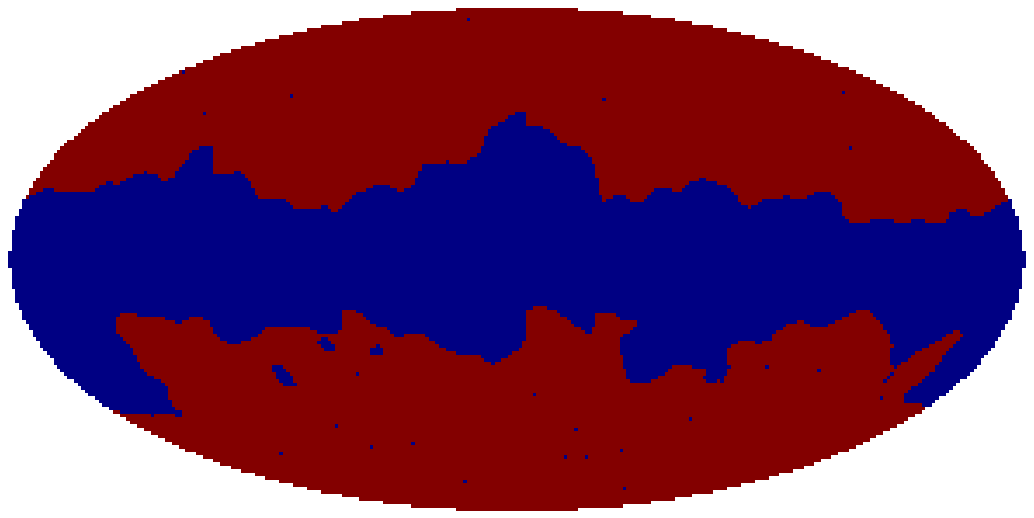}
  \end{center}
  \caption{Eigendecomposition of the CR system using a diagonal
    preconditioner.  {\em Top panel}: The eigenvalues of
    $\text{diag}(\A)^{-1}\A$ for the 143 GHz Planck channel
    with a mask covering 40\% of the sky, smoothed with a 5.6$^\circ$
    FWHM beam and truncated at $\lmax=95$. {\em Bottom panel:} A
    selection of eigenvectors corresponding to very low eigenvalues.
    The structure of the mask (bottom) is clearly visible in the
    eigenvectors.}
  \label{fig:eigen}
\end{figure}

As stressed in Section \ref{sec:sht}, $\Yobs^T$ denotes spherical
harmonic {\em adjoint synthesis}, and not spherical harmonic analysis. 
Pixels that are masked out, typically due to strong foreground
contamination, are simply missing from the data vector $\d$, and so
the corresponding rows are not present in $\Yobs$.  This means $\Yobs$
is not an orthogonal matrix, but that is not a concern since we never
perform spherical harmonic analysis of pixels on the observation grid.
The solution $\x$ is still well-defined everywhere on the sky due to
the prior term $\S^{-1}$.  This is typically implemented by
introducing zeroes in $\N^{-1}$ rather than removing rows of $\Yobs$,
which has the statistical interpretation of giving those pixels
infinite variance. The two interpretations are algebraically
equivalent.

\subsection{Eigenspectrum and CG performance}
\label{sec:conditioning}

The CR system in Equation \eqref{eq:system-explicit} is
symmetric and positive definite, which suggests the use of the
Conjugate Gradient (CG) algorithm. For the behavior of CG and other
Krylov methods, we are primarily interested in the eigenspectrum after
preconditioning \citep[and references therein]{shewchuk:1994}, i.e.
the eigenspectrum of $\mathbf{M}\mathbf{A}$, where
$\mathbf{M}\approx\mathbf{A}^{-1}$.  To illustrate the fundamental
problem with the CG algorithm for the application considered here, we
show in Figure \ref{fig:eigen} the eigenspectrum of a low-resolution
setup, using a diagonal preconditioner. This case corresponds to a
simulation of the 143 GHz Planck frequency map \citep{planck:2013a},
downgraded to an angular resolution of $5.4^{\circ}$, bandwidth-limited 
at $\ell_{\textrm{max}}=95$, and with a mask applied that
removes 40\% of the sky. The overall shape of the spectrum appears to
be mostly independent of the resolution, with a significant fraction
of degrees of freedom found in the tails. This behavior is 
representative of that found in real-world cases.

The problematic feature is the exponential drop in the eigenvalues
seen to the left of the figure. Theoretical results indicate that the
CG search needs at least one iteration per eigenvalue located in
exponentially increasing parts of the eigenspectrum
\citep{axelsson-lindskog-a,axelsson-lindskog-b}.  This leads to
extreme degradation of CG performance, which is indeed
what has been observed with Commander on high-resolution, 
high-sensitivity data.

The exponential spectral feature is due to large-scale modes under
the mask. For all but the smallest angular scales, the $\N^{-1}$ term dominates
by many orders of magnitude, so that the $\S^{-1}$ term is hardly seen
at all.  However, vectors that only build-up signal under the mask
after beam-smoothing will only see the $\S^{-1}$ term of the matrix,
as the $\N^{-1}$ term vanishes in that case. The eigenvectors
corresponding to the smallest eigenvalues are therefore characterized 
by having large scales localized within the mask. Moreover, the solution under
the mask is constrained by the values at the mask edge,
meaning the $\N^{-1}$ term takes effect, and this constraint is harder
closer to the edges. The result is an exponentially-falling eigenspectrum,
rather than separated clusters of eigenvalues that CG could more easily
deal with.

Phrased differently, for data having a high signal-to-noise ratio, the
pixels near the edge of the mask carry a large predictive power on the
signal inside the mask --- a signal that must be reconstructed by the CG
algorithm by navigating through a nearly degenerate system. In total,
the CG convergence rate is determined by a combination of the overall
signal-to-noise ratio and the size and shape of the mask. We 
have been unable to achieve proper convergence with this method for
the signal-to-noise ratio of a Planck-like experiment, for example, independent 
of preconditioners or number of iterations; downgrading and adding
regularization noise is required to produce robust results.

\section{The multi-level solver}

\begin{figure}
  \includegraphics[width=\linewidth]{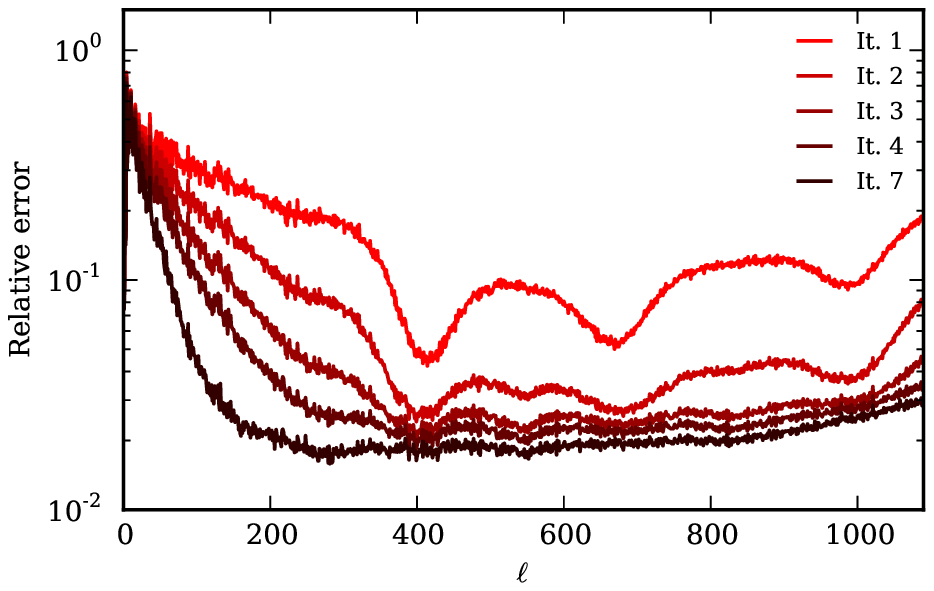}
  \setlength{\fboxsep}{0pt}%
  \setlength{\fboxrule}{1pt}%
  \begin{center}
    \fbox{\includegraphics[width=.45\linewidth]{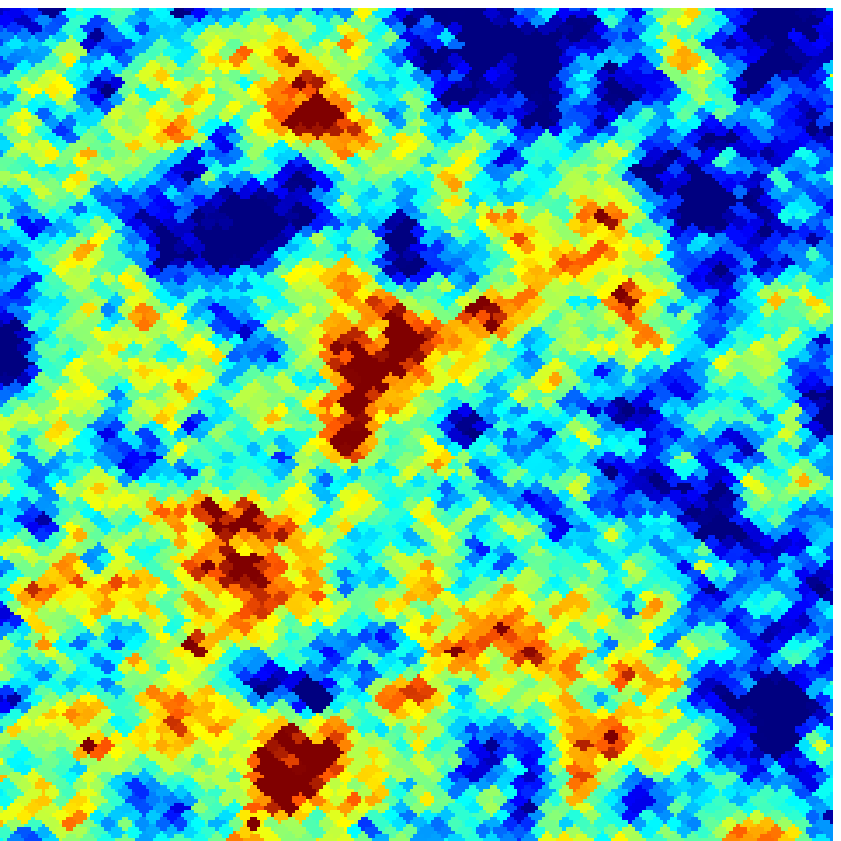}}
    \fbox{\includegraphics[width=.45\linewidth]{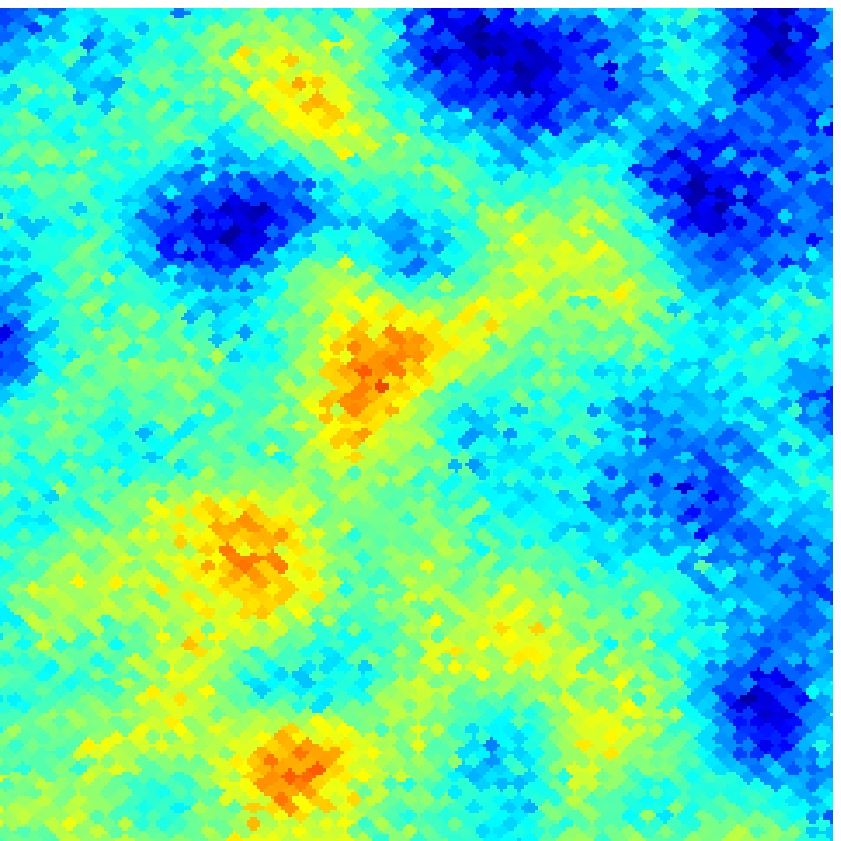}}\\
  \end{center}
  \caption{Effect of the error
    smoother/approximate inverse $\widehat{\M}$. {\em Top:} Relative
    error $\| \x_\ell - \x_{\text{true},\ell}
    \|/\|\x_{\text{true},\ell}\|$.  For each iteration, the error
    smoother developed in Section \ref{sec:smoother} is applied on a
    HEALPix $\Nside=512$ grid.  The error smoother is only able to get
    closer to the solution for some part of the frequency spectrum, and
    quickly stagnates since no improvement is made to the larger or
    smaller scales.  {\em Bottom:} The left patch shows the initial
    error when starting at $\x=\ve{0}$, while the right patch shows
    the error after the first iteration. The remaining large scale
    errors can be represented on a coarser grid.  This observation
    leads to the multi-level algorithm.}
  \label{fig:smoother}
\end{figure}

\subsection{Motivation for a multi-level method}
\label{sec:mg-overview}

The matrix $\A$ of Equation \eqref{eq:system-explicit} is defined in
spherical harmonic domain, and describes the coupling strength between
pairs of $(\ell, m)$ and $(\ell', m')$. Except in unrealistic
scenarios with very simple instrumental noise and mask, we have found
no pattern in the magnitudes of the matrix coefficients $A_{\ell
  m,\ell' m'}$ that is consistent enough to be exploited in a
solver.

By moving to pixel domain, however, we can {\it create} such an exploitable
pattern in the magnitudes of the matrix coefficients.  In Section
\ref{sec:localization} we will construct a corresponding pixel-domain
matrix $\widehat{\A}$ that is {\em localized}, in the sense that
$\widehat{A}_{ij}$ has small magnitude (less than $1\%$ of
$\widehat{A}_{ii}$) unless pixels $i$ and $j$ are very close together
on the sphere.

It is no surprise that the $\N^{-1}$ term of Equation
\eqref{eq:system-explicit} enjoys this property, since we have assumed
that instrumental noise is uncorrelated between pixels. When it comes
to the $\S^{-1}$ term, we note that $1/C_\ell$ is roughly proportional
to $\ell (\ell + 1)$, at least for $\ell \lesssim 1000$.  These are
the eigenvalues of the Laplacian on the sphere, with $\Y$ being the
corresponding eigenbasis. Therefore we can hope that a projection of
$\S^{-1}$ to pixel domain should be close to a Laplacian.  The
Laplacian is often approximated with a matrix where
$\widehat{A}_{ij}=0$ unless pixel $i$ and $j$ are neighbors or
$i=j$. While our case will be less perfect, it still suggests that
multi-level methods can be very efficient, since those are highly
successful for PDEs involving the Laplacian.

In Section \ref{sec:smoother}, we exploit the localization properties
in pixel domain to develop an approximate inverse
$\widehat{\M} \approx \widehat{\A}^{-1}$. Figure
\ref{fig:smoother} demonstrates the use of this approximate
solver as part of a simple stationary method
\begin{equation}
  \label{eq:smoothing-iterations}
  \x \leftarrow \x + \widehat{\M}(\b - \widehat{\A} \x),
\end{equation}
where we initialize $\x \leftarrow \ve{0}$ and then iteratively update
the solution. Note that if we replace $\widehat{\M}$ with
$\text{diag}(\A)^{-1}$, Eq. \eqref{eq:smoothing-iterations}
represents what are known as Jacobi iterations.

The problem that is evident from Figure \ref{fig:smoother} is that
$\widehat{\M}$ will only make improvements to one part of the
frequency spectrum --- namely, the highest frequencies that can be
represented on the grid used. This is the typical case when
multi-level methods are applied; iterations of the form of Equation
\eqref{eq:smoothing-iterations} are usually only efficient at
resolving the relations between pixels/elements that are strongly
coupled, which, when $\widehat{\A}$ is localized, translates to
resolving the solution at highest frequencies. Little or no
improvement is made between pixels that are weakly or indirectly
coupled in $\widehat{\A}$, so that no improvement is made to the
coarser scales.  Put another way, the error, $\ve{e} \equiv \x -
\x_\text{true}$, has its high-frequency components reduced, while the
low frequencies are left relatively unaffected. The approximate 
inverse $\widehat{\M}$ is therefore dubbed a {\em smoother} in
multi-level terminology. We will use the term {\em error smoother} to
distinguish it from the act of applying a low-pass filter (which is
instead called {\em restriction} in this context).

The key is now to project the matrix $\A$ to pixel grids at different
resolutions, producing a set of matrices $\widehat{\A}_h$, where $h$ is a level
indicator. For each $\widehat{\A}_h$ we construct a corresponding
error smoother $\widehat{\M}_h \approx \widehat{\A}_h^{-1}$ that resolves
the errors in one region of the frequency spectrum only. Using these
levels together, we arrive at a method that converges very well over
the entire frequency spectrum.

\subsection{The multi-level algorithm}
\label{sec:mg}

\begin{figure}
  \includegraphics[width=\linewidth]{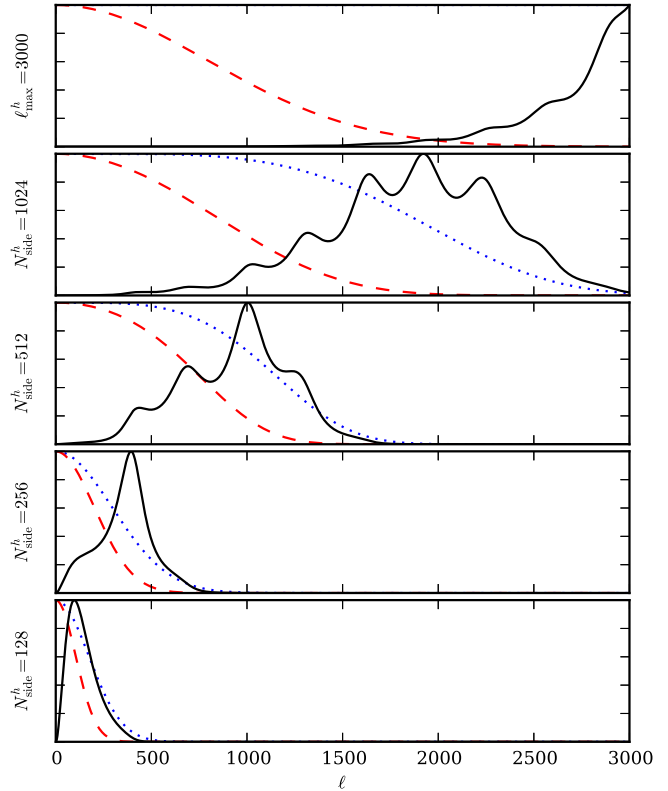}
  \caption{Effect of filters in harmonic domain for the top five
    levels. For each level $H$, starting from the original system of
    Equation \eqref{eq:system-explicit} at the top, we plot the transfer
    filter $f^H_{h,\ell}$ (dotted blue), the filtered prior
    $(\tilde{f}^H_\ell)^2/C_\ell$ (solid black), and an approximation
    to the diagonal of the inverse-noise term (dashed
    red). Functions are normalized to an arbitrary scale (see Figure
    \ref{fig:level-overview-ell-log} for the absolute scale). Note how
    the prior term on the pixel levels looks superficially similar to
    wavelets/needlets in harmonic domain \citep[and references
    therein]{scodeller:2011}. The real-space transform is also
    similar to wavelets/needlets (not plotted).}
  \label{fig:level-overview-ell}
\end{figure}

\begin{figure}
  \includegraphics[width=\linewidth]{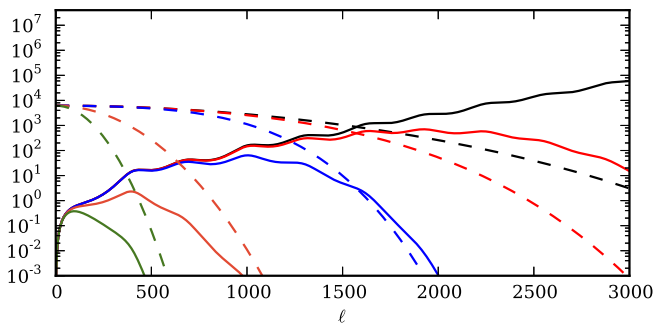}
  \caption{Same as Figure \ref{fig:level-overview-ell}, but all levels
    plotted together with a logarithmic scale and with absolute
    normalization. We plot the filtered prior
    $(\tilde{f}^H_\ell)^2/C_\ell$ (solid), and the diagonal of the
    inverse-noise term for 26 $\mu$K constant RMS and no
    mask (dashed). This noise level corresponds to the average of the
    RMS map of the 143 GHz Planck band. The levels are: The original
    system (black), $\Nsideh=1024$ (red), $\Nsideh=512$ (blue),
    $\Nsideh=256$ (orange), and $\Nsideh=128$ (green).  Note the effect
    of the filters on the signal-to-noise ratio; harmonic scales go
    from being data-dominated to noise-dominated at the point
    where the solid and dashed lines intersect.}
  \label{fig:level-overview-ell-log}
\end{figure}

In this section we give a brief overview of multi-level theory, together 
with the specification of our algorithm. For a more detailed introduction 
to multi-grid methods, consult one of the number of standard texts 
\cite[e.g.][]{hackbusch:1985}. Ingredients of multi-level algorithms are:

\begin{enumerate}
\item A set of bases to project the linear system into in order to work
  on different parts of the solution. Usually these form a hierarchy
  of levels from finest to coarsest, so that each level solves
  for different frequencies of the solution.  It is customary to
  label levels relatively, using $h$ for the current level and $H$
  for the coarser level.

\item A way to transfer vectors between the different levels.
  The {\em restriction} operator, $\I_h^H$, takes a vector from a
  finer level to a coarser level, while the {\em interpolation} operator
  $\I_H^h$ works in the opposite direction. For symmetric systems,
  one often takes $\I_H^h = (\I_h^H)^T$.

\item One linear operator (left-hand-side matrix) for each level. For 
  the case where interpolation is chosen to be transposed restriction, 
  these are often defined recursively as
  \begin{equation}
\label{eq:coarsen-system}
\A_H = \I_H^h \A_h (\I_H^h)^T
\end{equation}
 for the projection of a fine matrix $\A_h$ to a coarser matrix $\A_H$.

\item An {\em error smoother} $\M_h$ for each $\A_h$ that removes the
  higher frequencies of the error on level $h$, as discussed in the
  previous section.

\end{enumerate}

Multi-level algorithms are often implemented on a grid or a
tessellation in real space, with a sparse linear operator, and using
averages of neighboring points as the restriction operator
$\I_h^H$. In our case, $\widehat{\A}_h$ on each level is not sparse,
and, at least without approximations, multiplying $\widehat{\A}_H =
\I_H^h \widehat{\A}_h \I_h^H$ with a vector would be computationally
very expensive on the coarser levels as it would require interpolating
back to the highest-resolution grid.

To avoid this cost, we instead define our levels in spherical harmonic
domain.  Let $\tilde{f}_\ell^h$ be a spherical harmonic low-pass
filter that emphasizes one part of the frequency spectrum, and define 
$\F_h$ to be a diagonal matrix with elements $\tilde{f}_\ell^h$. We then define
\begin{equation}
  \label{eq:A_h}
  \A_h \equiv \F_h \A \F_h^T \equiv \D_h + \B_h \Yobs^T \N^{-1} \Yobs \B_h,
\end{equation}
where the prior term $\D_h$ is diagonal with entries given by
$(\tilde{f}^h_\ell)^2/C_\ell$ and the modified beam matrix $\B_h$ is
diagonal with elements given by $\tilde{f}^h_\ell b_\ell p_\ell$.
In this case, the system is bandlimited by some $\lmaxh \le \lmax$,
above which $\tilde{f}^h_\ell=0$.  Figures
\ref{fig:level-overview-ell} and \ref{fig:level-overview-ell-log} show
the filters used in our setup; we discuss the choice of
filters further in Section \ref{sec:localization}.

With this choice, we can clearly satisfy the multi-level hierarchy of
Equation \eqref{eq:coarsen-system} by choosing the restriction
operator $\I_h^H$ as an
$(\ell_\text{max}^H+1)^2$-by-$(\ell_\text{max}^h+1)^2$ block matrix,
where the block for $\ell \le \ell_\text{max}^H$ is diagonal with entries
\begin{equation}
  \label{eq:f_l}
  f_{h,\ell}^H \equiv \frac{\tilde{f}_\ell^H}{\tilde{f}_\ell^h},
\end{equation}
and the block for $\ell_\text{max}^H < \ell \le \ell_\text{max}^h$
is zero.

\newcommand{\find}{\quad}
\newcommand{\ind}{\quad \;}
\begin{figure}
  \begin{minipage}{.98\linewidth}

    \begin{tabular}{l}
      $\text{CR-Cycle}(h, \x, \b)$: \\
      \find {\bf Inputs:} \\
      \find  \ind $h$ -- The current level \\
      \find  \ind $\x$ -- Starting vector \\
      \find  \ind $\b$ -- Right-hand side \\
      \find \\
      \find \ind $H$ denotes the coarser level relative to $h$.\\
      \find {\bf Output:} \\
      \find \ind Improved solution vector $\x$ \\
      \\
    \end{tabular}
    \begin{tabular}{ll}
      \find {\bf if} $h$ is bottom level: & \\
      \find \ind $\x \leftarrow \A^{-1}_h \b$ & {\em By dense Cholesky}\\
      \find {\bf else}: \\
      \find \ind $\x \leftarrow \x + \Y_h^T \widehat{\M}_h \Y_h(\b - \A_h \x)$ & {\em Pre-smoothing} \\
      \find \ind $\r_H \leftarrow \I_h^H (\b - \A_h\x)$                & {\em Restricted residual} \\
      \find \ind $\ve{c}_H \leftarrow \ve{0}$               & {\em Coarse correction} \\
      \find \ind {\bf repeat} $n_\text{rec}^h$ times: & \\
      \find \ind \ind $\ve{c}_H \leftarrow \text{CR-Cycle}(H, \ve{c}_H, \r_H)$ & {\em Recurse }\\
      \find \ind $\x \leftarrow \x + (\I_h^H)^T \ve{c}_H$ & {\em Apply correction} \\
      \find \ind $\x \leftarrow \x + \Y_h^T \widehat{\M}_h \Y_h(\b - \A_h \x_)$ & {\em Post-smoothing} \\
      \find {\bf return} $\x$
    \end{tabular}
    \begin{tabular}{l}
      \\
      $\text{CR-Solve}(\b, \epsilon)$: \\
      \find {\bf Inputs:} \\
      \find  \ind $\b$ -- Right-hand side \\
      \find  \ind $\epsilon$\, -- Requested improvement in residual \\
      \find {\bf Output:} \\
      \find \ind Approximate solution $\x$\\
      \find \\
    \end{tabular}
    \begin{tabular}{ll}
      \find $\x \leftarrow \ve{0}$ & \\
      \find {\bf repeat:} & \\
      \find \ind $\x \leftarrow \text{CR-Cycle}(\text{1st}, \x, \b)$ & \\
      \find \ind $\r \leftarrow \b - \A\x$ & {\em Reused in next CR-Cycle} \\
      \find \ind {\bf if} $\r^T\S^{-1}\r < \epsilon \b^T \S^{-1} \b$: & 
      {\em Improvement relative to $C_\ell$} \\
      \find \ind \ind {\bf return} $\x$ & \\
    \end{tabular}
  \end{minipage}
  \caption{The multi-level CR solver. The matrices involved are
    defined in the main text.  In place of the simple iteration scheme
    of CR-Solve, one can use CR-Cycle as a preconditioner within
    another solver, such as CG.  By varying the $n_\text{rec}^h$
    parameter, a variety of solver cycles can be constructed, such as
    a V-cycle ($n_\text{rec}^h=1$) or W-cycle ($n_\text{rec}^h=2$).
    Note that, for simplicity, the top-level diagonal error correction
    is omitted; see the main text.}
  \label{code:mg}
\end{figure}

As already mentioned in Section \ref{sec:mg-overview}, the error
smoother that we have available, $\widehat{\M}_h$, is defined in pixel
domain. For every spherical harmonic (SH) level we therefore tag on a
corresponding sibling pixel level with matching HEALPix resolution
$\Nsideh$. The result is the following level structure:
\begin{align*}
  \text{SH at} & \text{ $\lmaxh=3000$} &\longleftrightarrow & \quad \text{Pixels at $\Nsideh=1024$} \\
  &\updownarrow & & \\
  \text{SH at} & \text{ $\lmaxh=2048$} &\longleftrightarrow & \quad \text{Pixels at $\Nsideh=512$} \\
  &\updownarrow & & \\
  \text{SH at} & \text{ $\lmaxh=1280$} &\longleftrightarrow & \quad \text{Pixels at $\Nsideh=256$} \\
  &\updownarrow & & \\
  &\;\vdots & &
\end{align*}
The arrows indicate that transfers between different scales 
happen only through the spherical harmonic levels. As emphasized
in Section \ref{sec:sht}, no spherical harmonic analysis operations
are performed in each conversion (only synthesis and adjoint synthesis
operations), and the implied under-pixelization in the above scheme is
therefore numerically unproblematic.

The full details of how to properly move between the levels to obtain
a solution is given in pseudo-code in Figure \ref{code:mg}. We
highlight some aspects in what follows.

Assume that we are currently on some
spherical harmonic level $h$ (where the original equation is simply
the top level), with corresponding system
\begin{equation}
  \label{eq:sh-level-system}
  \A_h \x_{\text{true},h} = \b_h.
\end{equation}
We start with some search vector $\x_h$ (initialized to zero), and want
to improve it to get closer to the true value $\x_{\text{true},h}$.
In order to make use of $\widehat{\M}$, we must now move to the
corresponding pixel level.  Our chosen restriction operator, denoted 
$\Y_h$, is spherical harmonic synthesis to a HEALPix grid\footnote{ 
  The pixel level is actually coarser than the spherical harmonic 
  level, because $\lmaxh$ must be chosen
  so high that the grid cannot resolve all the scales of the
  projected field. See Section \ref{sec:bandlimit}.} of resolution $\Nsideh$.
The key to efficient multi-level solvers is to transfer the residual vector
$\r_h$, and {\em not} the search vector $\x_h$;
\begin{align}
  \r_h &\leftarrow \b_h - \A_h \x_h\\
  \label{eq:pixel-correction-1}
  \widehat{\r}_h &\leftarrow \Y \r_h,
\end{align}
where $\widehat{\r}_h$ is the pixel domain projection of
$\r_h$. Then, we approximately solve the projected system
for a {\em correction vector} $\widehat{\ve{c}}_h$,
\begin{equation}
\label{eq:pixel-correction-2}
\widehat{\ve{c}}_h \leftarrow \widehat{\M}\, \widehat{\r}_h \approx
 (\Y_h \A_h \Y_h^T)^{-1}\, \widehat{\ve{r}}_h,
\end{equation}
where the computation of $\widehat{\M}\, \widehat{\r}_h$ is further
described in Section \ref{sec:smoother}. The approximation is better
for small scales than for large scales.  Finally, we let the
interpolation operator be the transpose of restriction, $\Y^T$, so that
the correction is brought over to the spherical harmonic search vector
by adjoint spherical harmonic synthesis,
\begin{equation}
\label{eq:pixel-correction-3}
  \x_h \leftarrow \x_h + \Y^T \widehat{\ve{c}}_h.
\end{equation}
Together, these steps act as the error smoothing of a spherical
harmonic level, labeled pre- and post-smoothing in Figure
\ref{code:mg}. Here we have motivated the procedure as arising from moving
between levels, but the idea of solving for a correction in a projected
system arises in many settings \citep{tang:2009}, and
other variations on this theme may prove fruitful in the future.

The vertical movement between coarser and finer levels follows the same
pattern, but uses the restriction operator $\I_h^H$ defined in
Equation \eqref{eq:f_l} instead of pixel projection $\Y$. First, a fine
residual is computed and restricted (i.e. low-pass filtered)
to the coarser level,
\begin{align}
  \r_h &\leftarrow \b_h - \A_h \x_h,\\
  \label{eq:coarse-correction-1}
  \r_H &\leftarrow \I^{H}_h \r_h.
\end{align}
Then, a coarse correction $\ve{c}_H$ is sought that approximates
the solution of the coarse system
 \begin{equation}
\label{eq:coarse-correction-2}
\I_h^H \A_h (\I_h^H)^T \ve{c}_H = \A_H \ve{c}_H = \r_H.
 \end{equation}
Except for at the bottom level, this happens by initializing a
search vector $\ve{c}_H$ to zero and recursively applying the algorithm.
Finally, the correction is interpolated and applied to our current 
search vector,
\begin{equation}
\label{eq:coarse-correction-3}
  \x_h \leftarrow \x_h + \I^{h}_H \ve{c}_H.
\end{equation}
Using this idea of transferring residuals and corrections between
levels with different bases, one can form a variety of multi-level {\em
  cycles}, moving between the levels in different patterns. Our choice
in the end is a W-cycle on the coarser levels and a V-cycle on the
finer levels, as described in Section \ref{sec:results} and the
pseudo-code.

In addition to the pixel levels described above, the top and bottom
levels are special. The smallest scales ($\ell \gtrsim 2200$ in
our experimental setup) are strongly noise-dominated, making the
spherical harmonic domain matrix $\A$ nearly diagonal. As a result, we do
not project to a pixel grid, but simply use $\text{diag}(\A)^{-1}$ as
the error smoother. Note, however, that this process would destroy the 
solution on scales that are not entirely noise-dominated, and so we first 
apply a high-pass filter to the correction vector before applying it to the
solution search vector. For the largest scales ($\ell \le 40$), we
do not project to pixel domain either, but simply solve $\A \x = \b$
restricted to $\ell \le 40$ by explicitly computing the matrix entries
and using a simple Cholesky solver. 

For the top solver level, we need to compute the diagonal of $\Yobs^T
\N^{-1} \Yobs$ in spherical harmonic domain, and for the bottom solver
level we similarly need all entries of $\Yobs^T \N^{-1} \Yobs$ for
$\ell$ up to some $\ell_\text{dense}$. While such entries can be
computed using Wigner 3j-symbols \citep{hivon:2002,eriksen:2004}, the
following procedure has some significant advantages. Firstly, while the
computational scaling is the same, it is much faster in practice, in
particular due to the optimized code for associated Legendre
polynomials $P_{\ell m}$ available in \verb@libpsht@ \citep{libpsht}.
Secondly, it is accurate to almost machine precision for any grid,
whereas the method relying on Wigner 3j-symbols relies on
approximation by evaluation of an integral, and is therefore inaccurate
for low-resolution HEALPix grids.

Let $\xi_{kj}$ be the $j$-th of $J_k$ pixels on ring $k$ in the masked
inverse-noise map. One can then evaluate \small
\begin{align*}
  \label{eq:Ni-dense}
  (\Yobs^T \N^{-1} &\Yobs)_{\ell_1 m_1,\ell_2 m_2}
= \\\quad\quad\quad&=\sum_{k}\sum_{j=1}^{J_k} \xi_{kj}
Y_{\ell_1 m_1}(\theta_j,\phi_{kj}) Y^*_{\ell_2 m_2}(\theta_j,\phi_{kj})\\
\quad\quad\quad&= \sum_{k}
    \widetilde{P}_{\ell_1 m_1}(\cos \theta_k) \widetilde{P}_{\ell_2 m_2}(\cos \theta_k)
\sum_{j=1}^{J_k} \xi_{kj} e^{i(m_1 - m_2)\phi_{kj}},
\end{align*}\normalsize
where the normalized associated Legendre function is
\begin{equation}
\widetilde{P}_{\ell m}(\cos \theta) = \sqrt{\frac{2\ell + 1}{4 \pi}\frac{(\ell - m)!}{(\ell + m)!}} ~ P_{\ell m}(\cos \theta).
\end{equation}
The inner sum can be precomputed for each ring $k$ and every $(m_1 - m_2)$
by discrete Fourier transforms, allowing the evaluation of
matrix elements in $O(N_\text{ring}) = O(\lmax)$ time. In the case that 
we want a dense low-$\ell$ block, the procedure
to downgrade the inverse-noise operator from Section \ref{sec:bandlimit}
should be applied first, to reduce the computational cost
from $O(\ell_\text{dense}^2 \lmax)$ to
$O(\ell_\text{dense}^3)$.


\begin{figure}
  \begin{center}
    \setlength{\fboxsep}{0pt}%
    \setlength{\fboxrule}{1pt}%
    \begin{tabular}{cc}
    \fbox{\includegraphics[width=.45\linewidth]{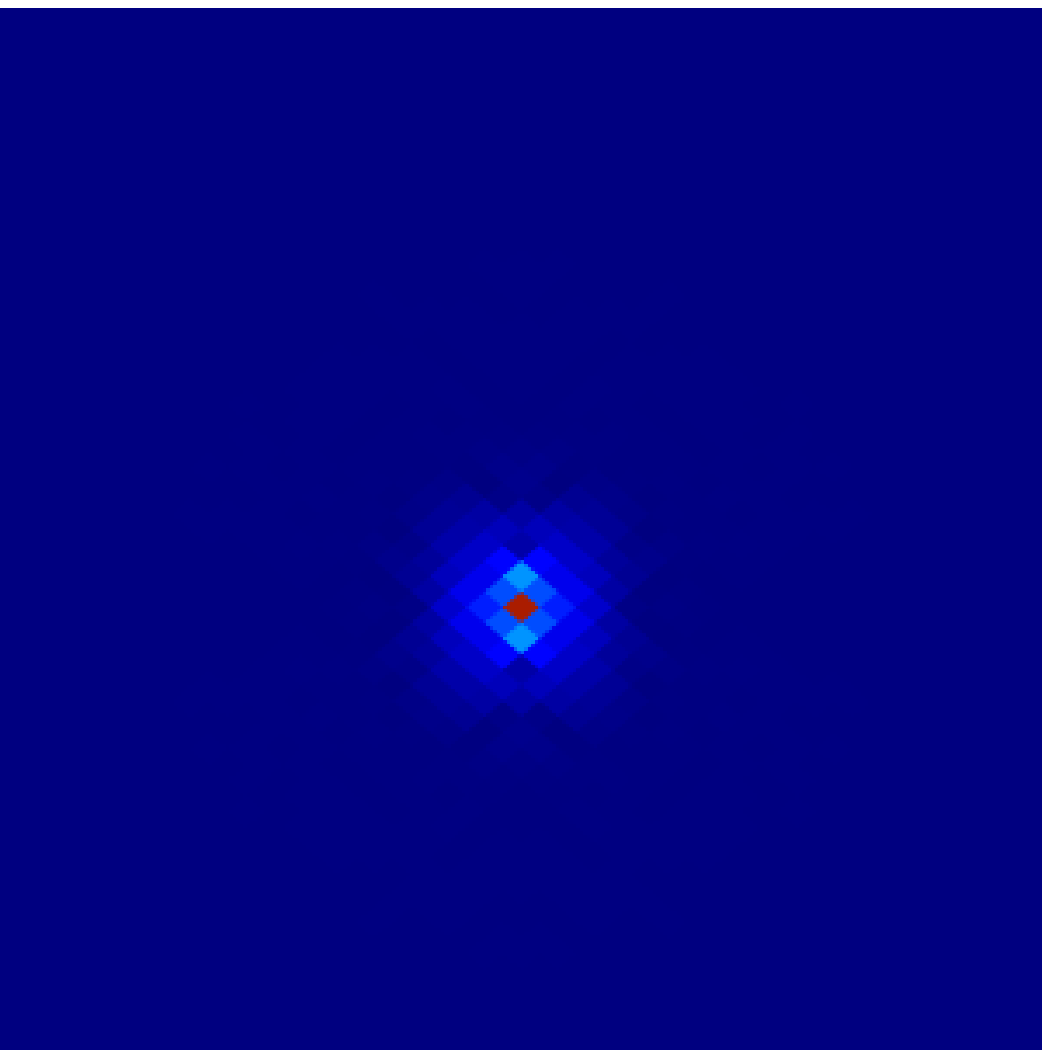}} &
    \fbox{\includegraphics[width=.45\linewidth]{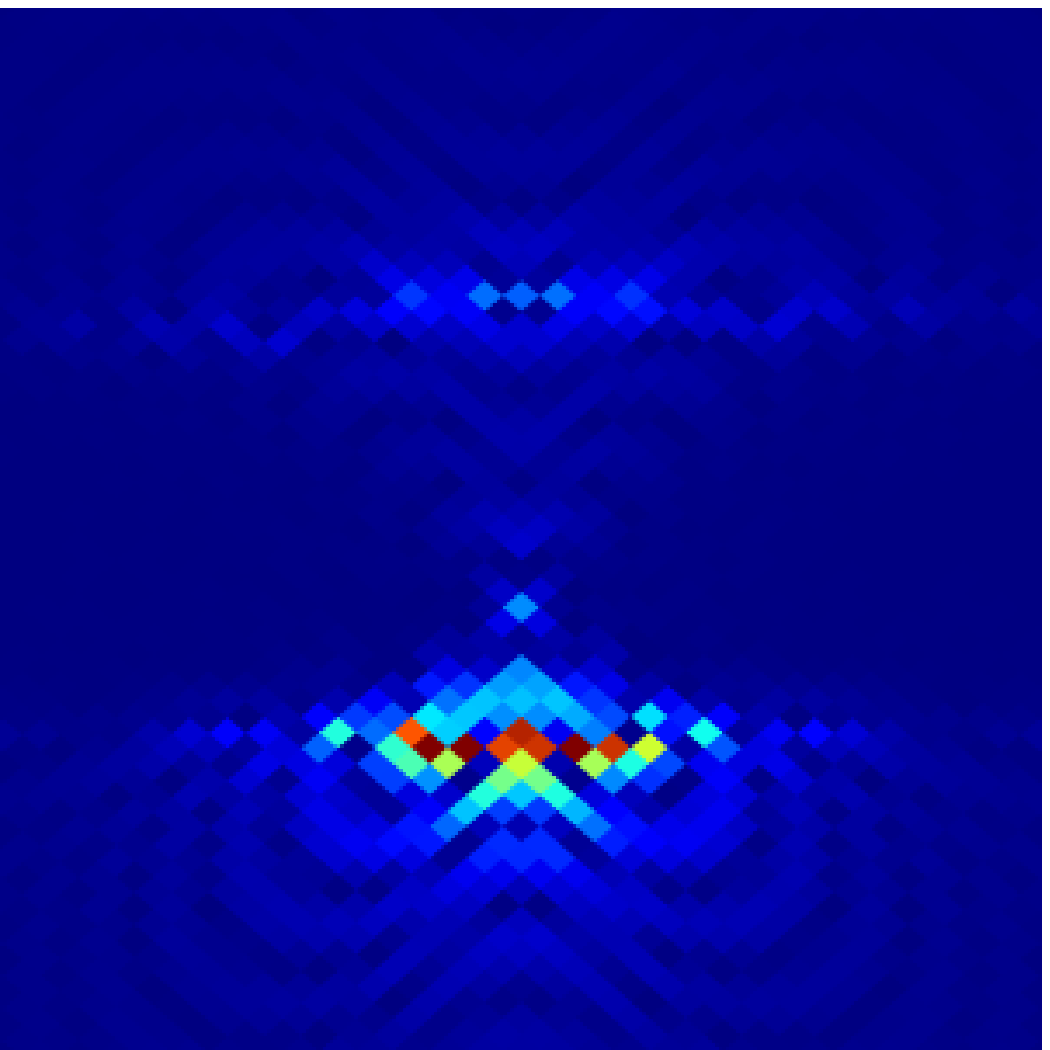}}\\[1mm]
      $\widehat{\D}_h$ & $\widehat{\B}^T_h \N^{-1} \widehat{\B}_h$ \\[3mm]
    \fbox{\includegraphics[width=.45\linewidth]{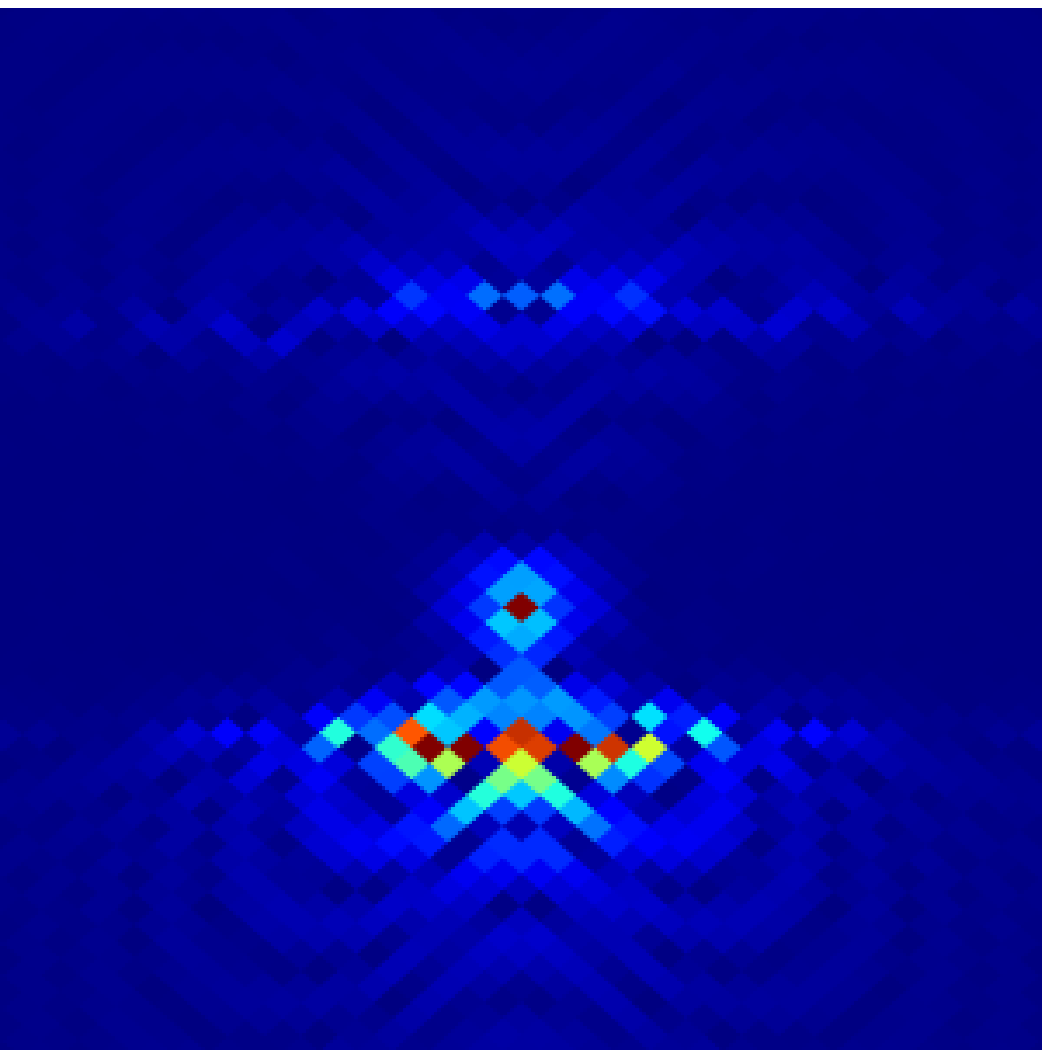}} &
    \fbox{\includegraphics[width=.45\linewidth]{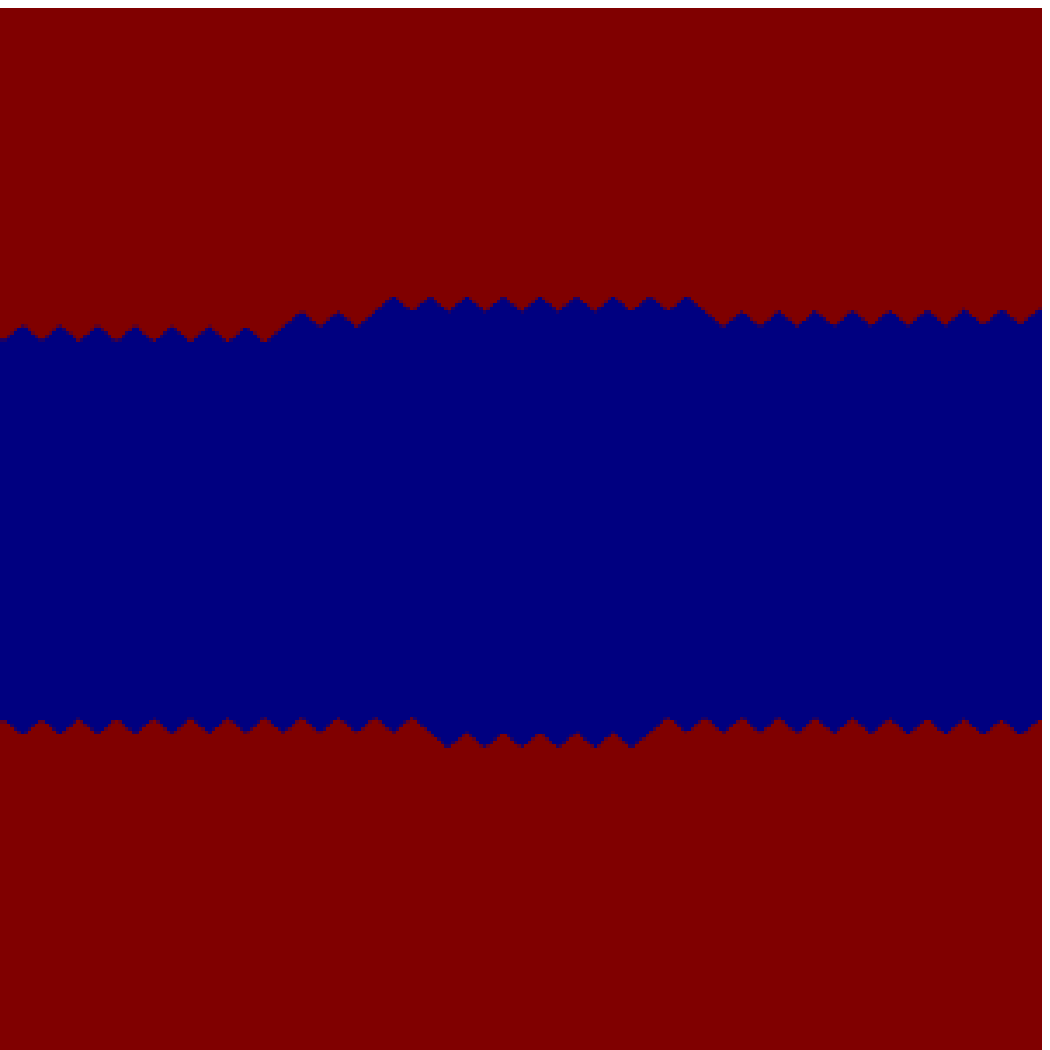}} \\[1mm]
      $\widehat{\A}_h = \widehat{\D}_h + \widehat{\B}^T_h \N^{-1} \widehat{\B}_h$ & {\small Mask} \\
    \end{tabular}
  \end{center}
  \caption{Effect of a poor choice of filter $\tilde{f}_\ell^h$.  Each
    panel shows the couplings between a single pixel and its
    neighboring region, corresponding to a row/column of
    $\widehat{\A}_h$.  In this case we used a low-pass filter based on
    modifying a standard needlet \citep[and references
    therein]{scodeller:2011}. While the harmonic properties of this
    filter were very attractive, the tails do not decay quickly enough
    in real space.  The resulting strong, long-range couplings are
    fatal to our algorithm.}
  \label{fig:ringing}
\end{figure}

\begin{figure*}
  \includegraphics[width=\linewidth]{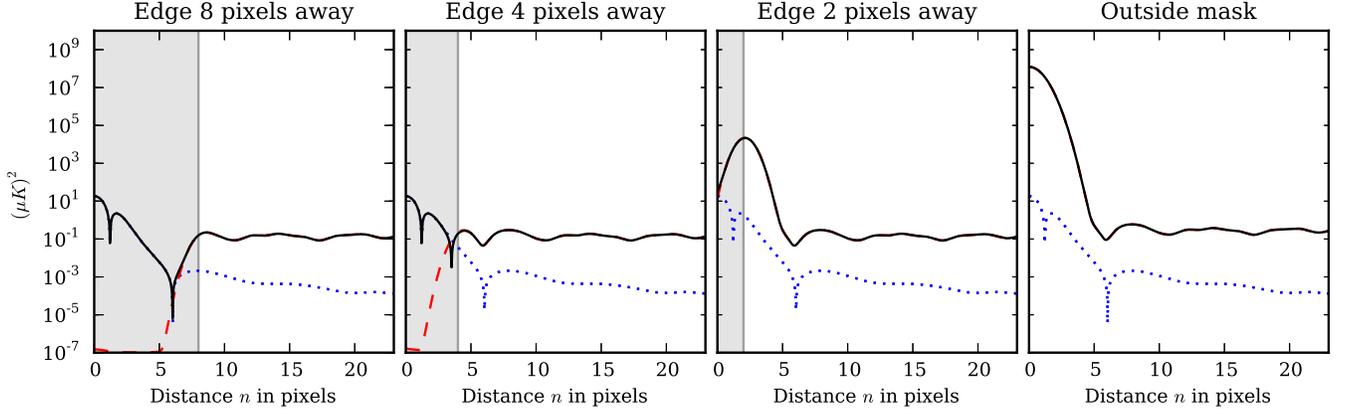}
  \caption{Effect of the mask on $\widehat{\A}_h$. Each panel shows
    the coupling strength in absolute value in the $\widehat{\A}_h$
    operator, between a sample point at $(\theta, \phi)$ (plotted at
    the origin), and another sample point $n$ pixels away at $(\theta,
    \phi + n\Delta)$, where $\Delta$ is the angular size of one pixel.
    The couplings of $\widehat{\A}_h$ (black) are a sum of the prior
    term $\widehat{\D}_h$ (dotted blue) and the inverse-noise term
    $\widehat{\B}_h^T \N^{-1} \widehat{\B}_h$ (dashed red).  For each
    panel, we vary the position of $(\theta, \phi)$ relative to the 
    mask (gray band), so that the origin is in
    each case a value on the diagonal of $\widehat{\A}_h$.  Displayed 
    here is our $\Nsideh=32$ level in the case of 1.9 $\mu$K constant
    RMS noise (the minimum RMS level of the Planck 143 GHz band). 
    The filter $\tilde{f}_\ell$ is a product of all the inter-level
    filters $f^h_{H,\ell}$ (as described in the text), but corresponds
    roughly to a Gaussian with FWHM of 2 pixels divided by the pixel
    window $p_\ell$. The ``floor'' at $10^{-1}$ is caused by the
    non-Gaussian features of the instrumental beam, $b_\ell$. For
    comparison, a perfect Gaussian instrumental beam is used in Figure
    \ref{fig:bandlimit}.}
  \label{fig:maskeffect}
\end{figure*}

\begin{figure*}
  \includegraphics[width=\linewidth]{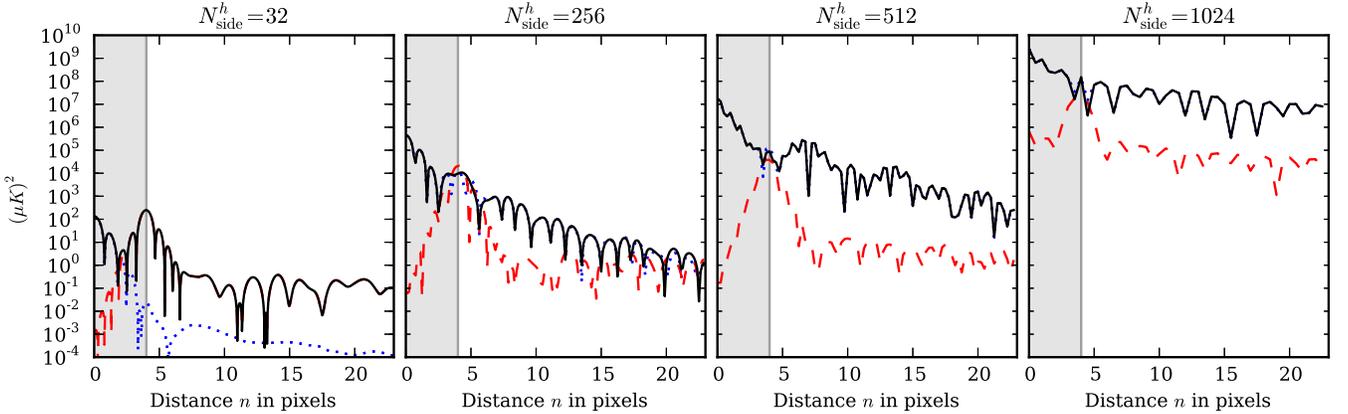}
  \caption{Effect of resolution on $\widehat{\A}_h$. See Figure
    \ref{fig:maskeffect} for legend and experimental setup. In this
    figure, we also show the effect of the filter $q_\ell$ of
    Equation \eqref{eq:fourth-power-beam}, with $\lambda$ appropriately
    tuned for the resolution in each case. As the resolution is
    increased, the signal-to-noise ratio decreases, making the
    influence of the edge of the mask less important.}
  \label{fig:resolution-effect}
\end{figure*}

\begin{figure*}
  \includegraphics[width=\linewidth]{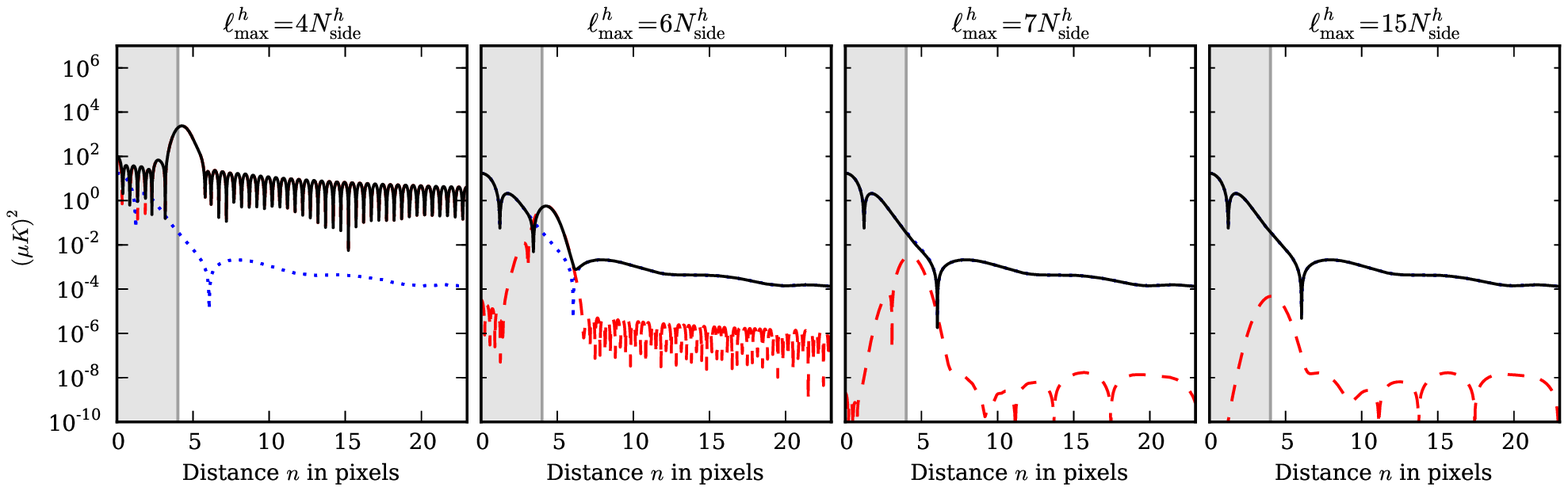}
  \caption{Effect of the band-limit $\lmaxh$ on $\widehat{\A}_h$. See
    Figure \ref{fig:maskeffect} for legend and experimental
    setup. The settings for each panel are the same except for varying 
    $\lmaxh$.  Here, the product $\tilde{f}_\ell^h b_\ell
    p_\ell$ is a pure Gaussian with FWHM of 2 pixels. Since the
    instrumental beam is in this case taken to be a perfect Gaussian,
    there is also no ``floor'' at $10^{-1}$ (compare with Figure
    \ref{fig:maskeffect} for the effect of a non-Gaussian beam).}
  \label{fig:bandlimit}
\end{figure*}

\subsection{Filter selection and pixel-domain localization}
\label{sec:localization}

So far we have not specified the exact form of the low-pass filters
$\tilde{f}_\ell^h$ required for every level. It turns out that careful
selection of these filters is essential to ensure that the pixel
projection of $\A_h$ is localized, and hence that the construction of
an efficient error smoother is possible.

As indicated in Eq. \eqref{eq:pixel-correction-2}, the
spherical harmonic system $\A_h$ on each level $h$ is projected
to pixel domain with
\begin{equation}
  \label{eq:pixel-domain}
 \widehat{\A}_h \equiv \Y_h \A_h \Y_h^T = \widehat{\D}_h + \widehat{\B}_h^T \N^{-1} \widehat{\B}_h,
\end{equation}
where the prior and pixelized beam terms are this time given by (respectively)
\begin{align}
 \label{eq:Dhat}
 \widehat{\D}_h &= \Y_h \F_h \S^{-1} \F_h \Y_h^T \\
 \label{eq:Bhat}
 \widehat{\B}_h &= \Yobs \B\; \F_h \Y_h^T.
\end{align}
Note that the pixelization along the rows of $\widehat{\B}_h$ is the
observational grid, while the pixelization down the columns is
that of the current level.

The matrices $\widehat{\D}_h$ and $\widehat{\B}_h$ are both
rotationally invariant. By the addition theorem of
spherical harmonics, the coupling strength between two points on the
sphere separated by angular distance $\theta$ is given by
\begin{align}
\label{eq:addition-theorem}
g(\theta) = \sum_\ell \frac{2\ell + 1}{4\pi} g_\ell P_\ell(\cos \theta),
\end{align}
where we insert $g_\ell = (\tilde{f}^h_\ell)^2/C_\ell$ for
$\widehat{\D}_h$ and $g_\ell = \tilde{f}^h_\ell b_\ell p_\ell$ for
$\widehat{\B}_h$. The pixel-domain localization of such matrices
depends entirely on $g_\ell$. In our experience, the $g_\ell$ that lead to
localized matrices in pixel domain tend to be flat or polynomially increasing
before an exponential drop. Since $b_\ell$ already describes a
localized beam, and $1/C_\ell$ increases non-exponentially, crafting a
localized system $\widehat{\A}_h$ at each level is indeed possible.

Selecting the filters $f^H_{h,\ell}$, whose products form
$\tilde{f}^h_\ell$ and $\F_h$ for each level, is a non-trivial
matter. The main characteristic the filters must have is that each
$\tilde{f}^h_\ell$ falls off quickly enough in real space to
avoid strong couplings between the edge of the mask and the
interior. Figure \ref{fig:ringing} shows what happens if this is not
controlled correctly --- the long-range couplings make the construction
of an error smoother $\widehat{\M}$ impossible. In contrast, Figure
\ref{fig:maskeffect} shows the behavior of the operators in the
well-tuned case.

A filter that we found to work very well is given by squaring the exponent
of a Gaussian,
\begin{equation}
\label{eq:fourth-power-beam}
q_\ell = \exp(-\ell^2 (\ell + 1)^2 \lambda).
\end{equation}
The scale parameter $\lambda$ is simply chosen from the scale
behavior that we want. In our test runs, we chose the constraints $q_{2570}
= 0.1$ at the $\Nsideh \!\!=\!\! 1024$ level and $q_{1536} = 0.1$ at the
$\Nsideh=512$ level.

\vspace{0.2em}
This filter has the following advantages over a simple Gaussian:
\begin{itemize}
\item It decays much more quickly in $\ell$, while in real space it
  decays almost as quickly in the tails as the Gaussian. This allows
  us to avoid increasing the bandlimit of the original system beyond
  $\lmax=3000$.
\item The rapid decay with $\ell$ is also beneficial to counter the
  behavior of $1/C_\ell$. In the range $2000 < \ell < 3000$,
  $1/C_\ell$ follows a rather steep trajectory (between $\sim\ell^7$ and
  $\ell^8$) which, when only countered by a Gaussian, causes some
  ringing and less locality.
\item Using Gaussian filters shapes the $\N^{-1}$ term so that
  couplings around a given pixel are similar to a Gaussian with FWHM
  of 4 pixels. That is, the couplings between neighboring pixels are
  rather strong. The filter defined above produces much weaker
  couplings between neighbors. This is not currently an advantage,
  because we let every pixel ``see'' a radius of $k=8$ pixels around
  itself anyway in the error smoother. However, it could become an
  advantage in the future if $k$ is chosen adaptively for each pixel.
\end{itemize}

Despite these features, the simple Gaussian filter behaved better at 
the coarser levels with very high signal-to-noise, as can be seen by 
comparing the second panel of Figure \ref{fig:maskeffect} with the 
first panel of Figure \ref{fig:resolution-effect}. 
In our tests we chose a Gaussian filter
$f_{h,\ell}^H$ for levels $\Nsideh \le 256$, tuned so that the
cumulative filter $\tilde{f}_\ell^h$ on each level roughly
corresponds to a Gaussian with FWHM of 2 pixels.

\subsection{Band-limitation and coarsening $\Y^T \N^{-1} \Y$}
\label{sec:bandlimit}

Figure \ref{fig:bandlimit} shows the effect of
choosing the bandlimit $\lmaxh$ too low. On the coarser levels,
ringing from the inverse-noise term causes strong non-local couplings
unless the bandlimit is set as high as $6 \Nsideh$. This limit depends
on the signal-to-noise ratio, and $\lmax=4\Nsideh$ is sufficient
on the $\Nside=512$ level.

The HEALPix grid can only represent a field accurately up to
$\lmaxh \sim 2\Nsideh$, and will in fact see different scales on
different parts of the sphere, due to the necessary irregularities in
the pixelization. This is the primary reason for the non-traditional
level traversal structure chosen in Section \ref{sec:mg}. The pixel
projection operator $\Y_h$ removes some parts of the projected field
that the grid cannot represent, but this is after all how
a multi-level restriction normally works, and so poses no problems.

The filter $\tilde{f}_\ell^h$ allows us to set $\lmaxh$ much lower
than the full $\lmax$. The two SHTs involved in $\Yobs^T
\N^{-1} \Yobs$ still involve an $\Nside=2048$ grid, however, so the
coarsest levels are still almost as computationally expensive as the
finest levels.

To work around this, the key is to note that the operator $\Yobs^T \N^{-1}
\Yobs$ does not ``see'' scales in the inverse-noise map
beyond $2\lmaxh$. This follows from an expansion into Wigner
3j-symbols \citep{hivon:2002,eriksen:2004}.  Simply degrading the
inverse-noise map to a coarser resolution HEALPix grid was found to be
far too inaccurate, so more care is needed. First, we rewrite the 
operator as
\begin{equation}
  \label{eq:Ni-rewrite}
  \Yobs^T \N^{-1} \Yobs = \Yobs^T \W_\text{obs} (\W_\text{obs}^{-1} \N^{-1}) \Yobs,
\end{equation}
where $\W_\text{obs}$ denotes the quadrature weights of the HEALPix
$\Nside=2048$ grid, so that $\Yobs^T \W_\text{obs}$ corresponds to
spherical harmonic analysis, as described in Section
\ref{sec:sht}. Then, we write $\xi_i$ for the pixels on the diagonal of
$\W^{-1} \N^{-1}$, and $\xi_{\ell m}$ for the same map expanded into
spherical harmonics. Since the operator of Equation
\eqref{eq:Ni-rewrite} does not see coefficients beyond $2\lmaxh$, we can
truncate $\xi_{\ell m}$ and project it onto a Gauss-Legendre grid of the
same order, which (unlike HEALPix grids) allows spherical harmonic
analysis that is accurate to almost machine precision. Using this
re-weighted and downgraded inverse-noise map as the diagonal of a new
inverse-noise matrix $\tilde{\N}^{-1}_h$, we have that
\begin{equation}
  \label{eq:Xi}
  \Yobs^T \N^{-1} \Yobs
= \tilde{\Y}^T_h \tilde{\ve{W}}_h \tilde{\N}^{-1}_h \tilde{\Y}_h,
\end{equation}
where $\tilde{\Y}_h$ and $\tilde{\Y}^T_h \tilde{\ve{W}}_h$ indicate
spherical harmonic synthesis and analysis on the Gauss-Legendre grid.

\subsection{Error smoother construction for the CR system}
\label{sec:smoother}

A simple diagonal error smoother does not converge in our setup,
primarily because pixels on the edge of the mask can have a very
strong influence on the solution in the interior of the mask, as seen
in Figure \ref{fig:maskeffect}. Also, when applying Gaussian filters,
the couplings between neighboring pixels are rather strong,
preventing the use of a diagonal error smoother even far from the
mask.

The basic strategy for our error smoother is to make sure that every
pixel ``sees'' neighboring pixels in some radius $k$ around it.  In
our case we let $k=8$ on all levels, although improvements on this may
be possible, especially in cases with lower signal-to-noise than ours.

\begin{figure}
  \begin{center}
    \includegraphics[width=.9\linewidth]{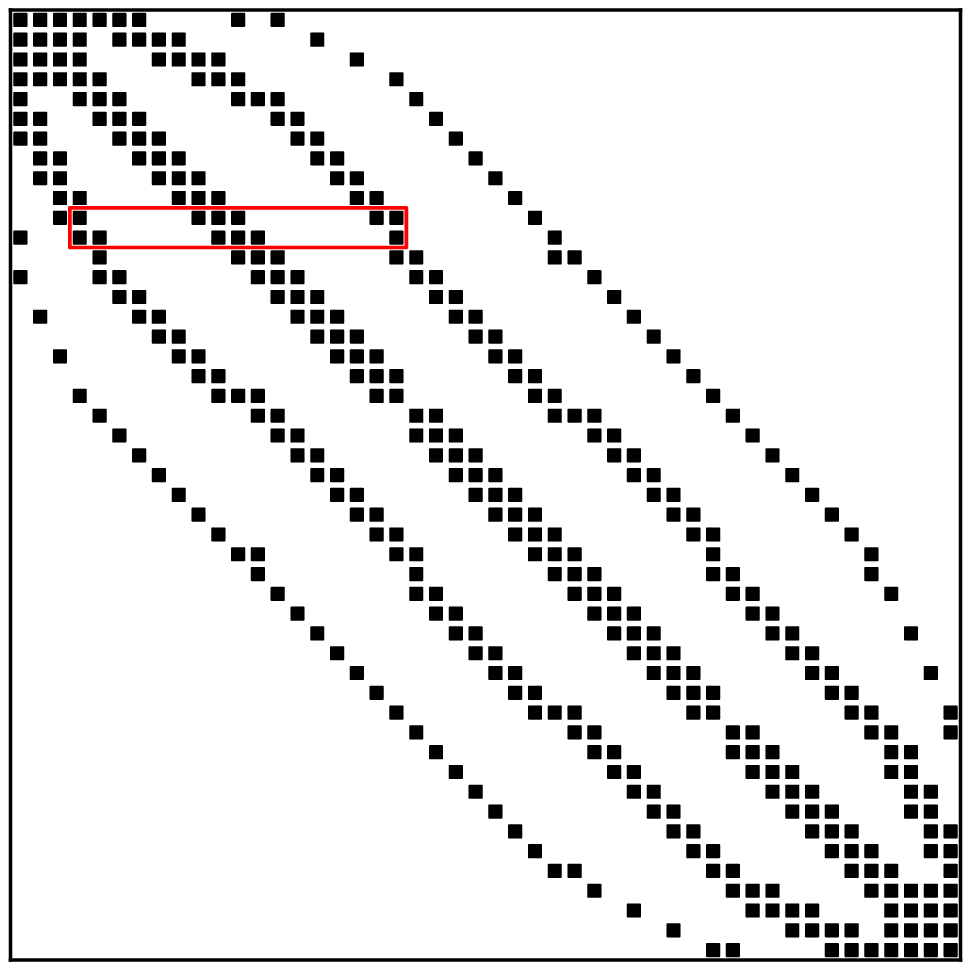}
    \includegraphics[width=.9\linewidth]{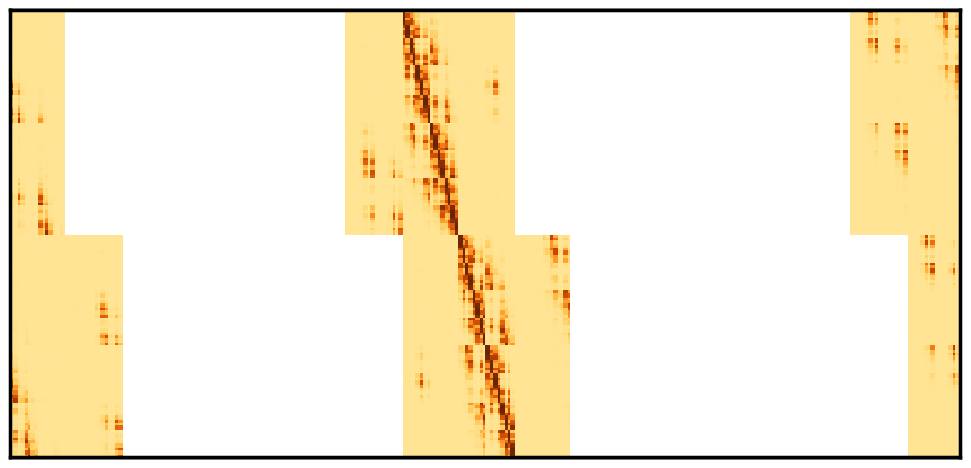}
  \end{center}
  \caption{Structure of the block sparse matrices used in the error
    smoothers. {\em Top panel}: The sparsity pattern when every tile
    is coupled to its 8 neighboring tiles. In this case, the pattern
    of tiles is an $\Nside=2$ HEALPix grid in ring-ordering.  {\em
      Bottom panel}: The blocks of $\widehat{\B}_h = \Yobs \B \Y_h^T$
    corresponding to the red rectangle in the top panel. The blocks on
    the diagonal contain within-tile couplings, while off-diagonal
    blocks are couplings between pixels in neighboring tiles. Each
    block is rectangular because $\Yobs$ samples on a grid with $4 \times$
    more pixels than the grid sampled by $\Y_h$.}
  \label{fig:smoother-structure}
\end{figure}

We start by dividing the sphere into tiles of size $k$-by-$k$. Then, we
include the couplings between pixels in the same and neighboring
tiles while ignoring any couplings between pixels further apart, so
that couplings are included in a radius of at least $k$ pixels around
every pixel. The result is a block sparse matrix, as shown in Fig.
\ref{fig:smoother-structure}.

Next, we explicitly compute the parts of $\widehat{\D}_h$ (Eq.
\eqref{eq:Dhat}) and $\widehat{\B}_h$ (Eq. \eqref{eq:Bhat}) that fall
within the sparsity pattern by evaluating the sum over Legendre
polynomials from Equation \eqref{eq:addition-theorem}.  After preparing
the block sparse matrix approximations, we use matrix multiplication
without fill-in to compute $\widehat{\B}^T \N^{-1} \widehat{\B}$ --- that
is, we neglect resulting blocks outside of the same sparsity
pattern. The approximant for $\widehat{\D}_h$ can then be added
directly.  Finally, we perform a zero-fill-in Incomplete Cholesky
factorization (ICC), i.e. we perform in-place Cholesky
factorization of the block sparse approximant as usual, but 
ignore any element updates outside of the sparsity pattern during the
factorization process.

Without modification, the factorization process usually fails, either
due to the sparse approximant of the full dense matrix ending up
non-positive-definite, or because of elements dropped during the
ICC. When this happens, we do a binary search for the lowest ridge
adjustment $\alpha$ that, when added to the diagonal, makes the factorization
procedure succeed, and scale this $\alpha$ by a factor of $1.5$ for
the final factorization. Typical ridge values $\alpha$ are in the
range $10^{-2}$ to $10^{-4}$ times the maximum element of $\A_h$.

After factorization, applying the smoother is simply a matter of doing
the usual triangular solve. This is an inherently sequential process,
and the smoother therefore currently runs on a single CPU core.  Since an error
smoother only needs to work locally, we expect to be able to apply
domain decomposition techniques, partitioning the sphere into large
domains that overlap by $k$ or $2k$ pixels, and applying one error
smoother on each domain. Proper parallelization of the error smoother
is left for future work, however. Also note that the process described 
above is the very simplest incomplete factorization algorithm, and more
sophisticated incomplete factorization algorithms are standard in the
literature.

In Section \ref{sec:results}, we quote numbers for the execution time and
memory usage of the smoother. One possibility for reducing memory consumption 
in the future is to let $k$ be adaptive, as it can be made smaller away
from the edges of the mask. All error smoother computations are done
in single precision. In the current implementation, computing $\widehat{\B}$
is very expensive, as we sample it directly on the $\Nside=2048$ grid.
This is not a fundamental scaling problem, but rather an issue of
implementation, as the degraded inverse-noise map
on the Gauss-Legendre grid described in Section \ref{sec:bandlimit}
could also be used in this setting.

\section{Implementation and results}

\subsection{Numerical results and performance}
\label{sec:results}

\begin{figure}
  \includegraphics[width=\linewidth]{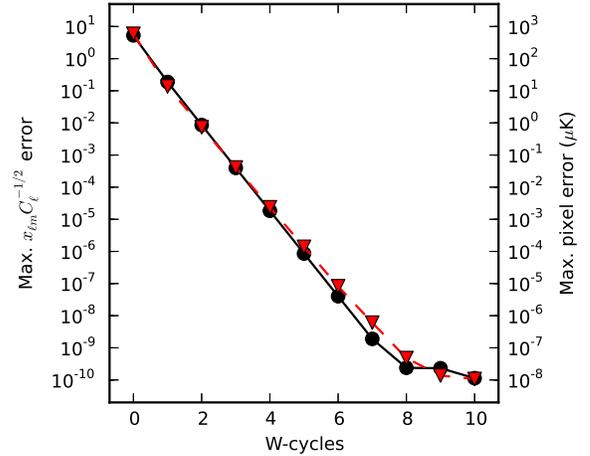}
  \caption{Absolute errors as a function of W-cycle count.  For every
    iteration we plot the maximum error over all $C_\ell^{-1/2}
    x_{\ell m}$ (black circles, left axis), as well as the largest
    error across all pixels (red triangles, right axis).}
  \label{fig:errors}
\end{figure}

\begin{figure}
    \includegraphics[width=\linewidth]{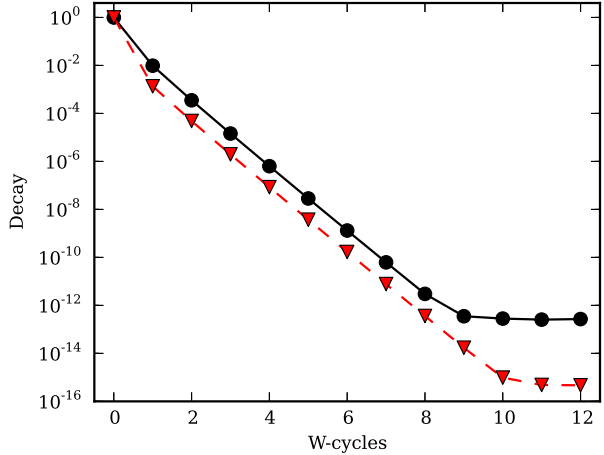}
  \caption{Comparison of absolute errors relative to $C_\ell$, $(\x -
    \x_\text{true})^T \S^{-1} (\x - \x_\text{true})$ (black circles),
    and similarly scaled residuals, $(\b - \A \x)^T\S^{-1}(\b -
    \A \x)$ (red triangles). Both are normalized with respect to the initial
    error/residual.  The two quantities behave very similarly, implying 
    that the residual is an excellent proxy for the true error.}
  \label{fig:residuals}
\end{figure}

\begin{deluxetable*}{lcccccc}
\tablecaption{\label{tab:cycle-cost}Structure and computational cost of a W-cycle}
\tablecomments{All times are given in wall time seconds
    using 16 CPU cores. The total number of operations of each kind
    for the W-cycle is indicated in each case; this number is not a multiple
    of the number of visits because the input vector $\x$ is zero on the first visit (except on
    the first level). Ignoring this aspect, each
    pixel level requires: i) two level-transfer spherical harmonic transforms ($\Y_h$),
    ii) three multiplications with $\A_h$, each with two inverse-noise
    spherical harmonic transforms ($\Yobs$), and iii) two applications of the
    error smoother $\widehat{\M}$. The top spherical harmonic level also
    requires two applications of $\A_h$, while the smoother application time is
    negligible. The bottom spherical harmonic level consists only of dense triangular
    solves.}
\tablecolumns{7}
\tablehead{ &  &  & Time in $\Y_h$ & Time in $\Yobs$
          & Time in $\widehat{\M}_h$ or $\A_h^{-1}$ & Total time \\
Level & $\lmax$ & \# of visits & (wall s) & (wall s) & (wall s) & (wall s) }
\startdata
      $\lmax=3000$    &  3000 & 1 &
          --- &
          6 $\times$ 2.8 &
          --- &
          16.8 \\

      $\Nside=1024$   &  3000 & 1 &
          4 $\times$ 1.2 &
          4 $\times$ 2.8 &
          2 $\times$ 11 &
          38.0 \\

      $\Nside=512$    & 2048 & 2 &
          8 $\times$ 0.3 &
          10 $\times$ 1.3 &
          4 $\times$ 2.3 &
          24.6 \\

      $\Nside=256$    & 1280 & 4 &
          16 $\times$ 0.07 &
          20 $\times$ 0.40 &
          8 $\times$ 0.57 &
          13.7 \\

      $\Nside=128$    & 768 & 8 &
          32 $\times$ 0.016 &
          40 $\times$ 0.10  &
          16 $\times$ 0.14 &
          6.75 \\

      $\Nside=64$     & 384 & 16 &
          64 $\times$ 0.004 &
          80 $\times$ 0.03  &
          32 $\times$ 0.035 &
          3.78 \\

      $\Nside=32$     & 224 & 32 &
          128 $\times$ 0.002 &
          160 $\times$ 0.008 &
          64 $\times$  0.009 &
          2.11 \\\vspace*{1mm}

      $\lmax=40$      & 40 & 32 &
          --- & --- &
          32 $\times$ 0.028 &
          0.90 \\\vspace*{1mm}
      
      Other work  &&&&&& 8 \\
      Full W-cycle &&&&&& 114 
\enddata
\end{deluxetable*}

The basic assumptions for our experimental setup have already been laid 
out in Section \ref{sec:data-model}. We choose for our example the RMS 
map and symmetric beam approximation of the 143 GHz channel of Planck, 
as provided in the Planck 2013 data release \citep{planck:2013a}.

We tried running both with the 40\%-sky, 80\%-sky and 97\%-sky masks
used in the Planck analysis, in all cases together with the 143 GHz
point source mask. The mask has some impact on speed of convergence,
but not enough to warrant attention, and we therefore only present the
results from the 80\%-sky mask, which was the {\em slowest} to converge.

For the power spectrum, $C_\ell$, we use the standard best-fit
Planck+WP+high-$\ell$ 6-parameter $\Lambda$CDM spectrum
\citep{planck:2013d}, but set $C_0$ and $C_1$ to the value of
$C_2$ as a wide prior for any residual monopole or dipole
component. Statistically, the prior for the monopole and dipole is of
little relevance, since the data so strongly constrain these
components. Note that the present algorithm will not let us condition 
on a given monopole and dipole (i.e. set $C_0 = C_1 = 0$), at least 
without modifications.

To produce the right-hand side, $\b$, corresponding to a random test
realization, we draw a simulated $\x_\text{true}$ from the prior $p(\s
| C_\ell)$, and multiply it with $\A$ of Equation
\eqref{eq:system-explicit}.  This synthetic setup allows us to track the
true error, $\e = \x_{\text{true}} - \x$. In a real setting the right
hand side is of course generated from observed data, and in this case
one can only track the residual, $\r = \b - \A\x$.

The error smoothers are least efficient on the largest scales.  At the
same time, these are much cheaper to process than the small-scale
smoothers due to the $O(\ell^3_\text{max})$ scaling of the spherical
harmonic transforms. We therefore choose a partial W-cycle, where the levels
for $\Nsideh \le 1024$ participate in a W-cycle ($n_\text{rec}^h=2$ in
Figure \ref{code:mg}), but the very expensive error smoother of the
$\Nsideh=1024$ level, as well as SHTs at $\lmax=3000$, are only run once
on the way down and once on the way up (a V-cycle).

In Figure \ref{fig:errors} we plot the resulting convergence, in terms
of absolute error as a function of W-cycle iteration count. Here we
see that the error falls exponentially with cycle count, at the rate
of roughly one order of magnitude per iteration. The largest error
anywhere on the sky is smaller than $1\mu\textrm{K}$ after only 3
W-cycles, and approaches the numerical precision limit after 8 cycles.

As mentioned above, since we know what the true solution is for the
simulated data, we are also able to trace the absolute error, $\e =
\x_{\text{true}} - \x$, although only the residual, $\r = \b - \A\x$, 
is available in real-world applications. Figure \ref{fig:residuals} shows 
that these have qualitatively very similar behavior as a function of 
W-cycle count, which implies that the residual can be used as a robust 
proxy for the actual error for the multi-level
algorithm. The same is not true for the CG method, for which the error
can flatten earlier than the residual due to the presence of the
nearly singular modes in $\A$.

\begin{figure*}
    \includegraphics[width=\linewidth]{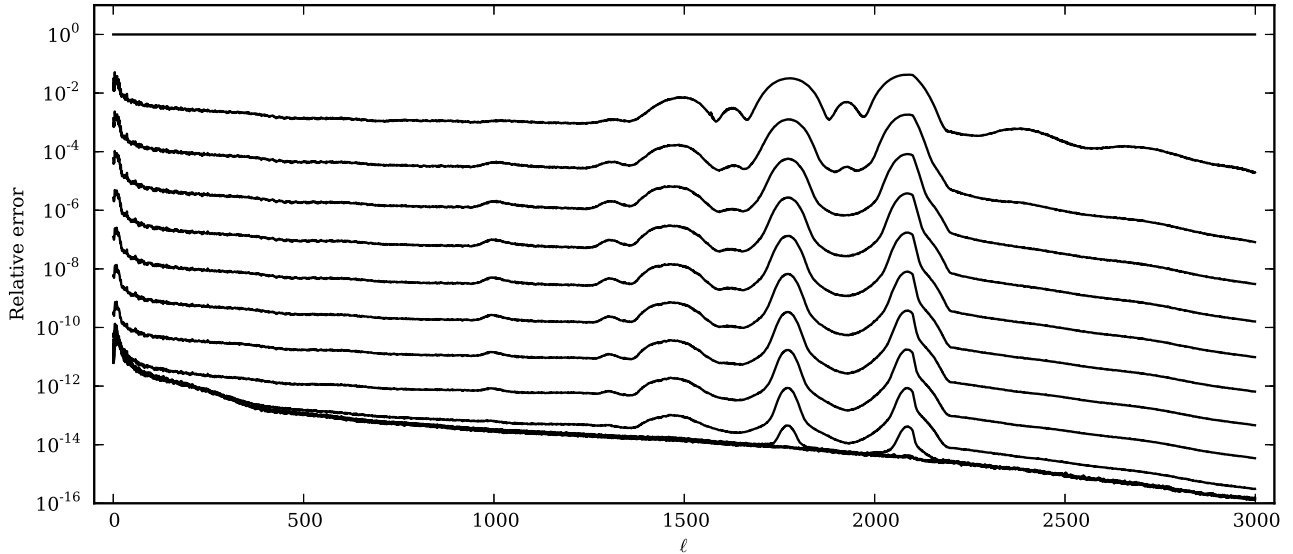}
  \caption{Relative error as a function of angular scale. Starting from 
  the top, each line shows the error for a given multi-level
    W-cycle.  Specifically, we plot $\| \x_{\text{true},\ell} -
    \x_\ell \| / \| \x_{\text{true},\ell} \|$, where $\x_\ell$ denotes
    a vector with the coefficients for a given $\ell$ only.  This plot
    is especially useful during development and tuning of the code, as one can
    immediately see which error smoothers do not perform well. \\}
  \label{fig:errors-by-scale}
\end{figure*}

Finally, in Figure \ref{fig:errors-by-scale} we show the relative
error as a function of multipole moment and W-cycle count. This plot
highlights the problematic angular scales, and is therefore
particularly useful during the debugging and tuning phase of the
analysis; for example, the use of a V-cycle rather than a
W-cycle would make the large scales noticeably lag behind in
convergence on this plot. Another example is that, if the filters
$\tilde{f}_h$ are poorly-tuned (potentially causing the method to
diverge), the responsible level can often be picked out on this plot.

The total run-time for this setup was 114 seconds wall time per W-cycle 
on 16 CPU cores (AMD 6282 running at 2.6 GHz). Table \ref{tab:cycle-cost} 
breaks this cost down further to the individual levels and actions. The bulk
of the memory use is by the error smoothers, which consume about 20 GiB
of memory (see Table \ref{tab:smoother-cost}). The total process
footprint was around 30 GiB, although unnecessary temporary arrays
abound in the current implementation.

Table \ref{tab:smoother-cost} presents the cost of the necessary
precomputations. For every new combination of instrumental beam, noise
map and mask, or for a new choice of multi-level filters
$f_{h,\ell}^H$, one needs to precompute an approximation to
$\widehat{\B}_h^T \N^{-1} \widehat{\B}_h$ for every solver level.
These precomputations required a total of 44 CPU hours in our tests, but
are trivially parallel. We also expect that one will usually load
the results from disk. The approximation for $\widehat{\D}_h$ must be
recomputed every time $C_\ell$ changes, which in the case of Gibbs
sampling means every time one wants to run the solver. Fortunately,
this computation is much cheaper and only requires around 100 CPU
minutes of trivially parallel work, plus 2 minutes of non-parallel work.
We argue in Section \ref{sec:discussion} that it should be possible to
greatly decrease precomputation time in future.

\begin{deluxetable}{rcccc}\vspace{-2em}
  \tablewidth{\linewidth}
  \tablecaption{\label{tab:smoother-cost}Error
    smoother precomputation cost/memory use per solver level}
  \tablecomments{ All times are given in CPU minutes (wall time times
    the number of CPU cores used). Precomputations can be divided into
    the part that must be performed whenever the observational setup
    (beam/mask/noise map) changes and the part that must be performed
    whenever the prior ($C_\ell$) changes. If any part changes, the
    non-parallel Incomplete Cholesky factorization (ICC) must also
    be performed again.} 
  \tablecolumns{5}
  \tablehead{ & Time obs. & Time $C_\ell$ & Time ICC & Mem. use \\
    $\Nside$ & (CPU min) & (CPU min) & (CPU min) & (GiB)} \startdata
  1024 & 727  & 85    & 1.15  & 15      \\
  512  & 509  & 15    & 0.35  &  3.7    \\
  256  & 340  &  2.4  & 0.10  &  0.93   \\
  128  & 230  &  0.36 & 0.02  &  0.23   \\
  64 & 452 & 0.05 & 0.007 & 0.058 \\\vspace*{1mm}
  32   & 363  &  0.01 & 0.002 &  0.014  \\
  Total & 2621 & 103 & 1.6 & 20
\enddata
\end{deluxetable}

The main weakness in the current implementation is the lack of
parallelization in the error smoothers. Not only does the code need to
be run on a single node, but the 40 seconds spent on error smoothing
runs on a single CPU core, with the 15 other cores
idling. Parallelization of the smoother would bring the wall time
much closer to 80 seconds, as well as allowing the distribution of the 
20 GiB of smoother data among several cluster nodes.

\subsection{Notes on implementation and dependencies}

The CR solver is part of Commander 2, which is made available as open 
source software under the BSD license (core code) and the GPL license 
(full software when including dependencies). For more information, see
\verb@http://commander.bitbucket.org/@.

Commander 2 is implemented in a mixture of Python (using NumPy and
SciPy), Cython \citep{cython}, Fortran 90, and C. For SHTs we use \verb@libsharp@
\citep{reinecke:2013}. For our benchmarks we have used OpenBLAS \citep{gotoblas,openblas} for
linear algebra.

The main computation time is spent in \verb@libsharp@ or OpenBLAS, and
as such is already highly optimized. The computation of Equation
\eqref{eq:addition-theorem} benefited greatly from being structured as
described in the appendix of \cite{seljebotn:2012}.  In addition to
what is mentioned there, we made use of the AVX and FMA4 instruction
sets. Also, note that all the computations for the error smoother could be
performed in single precision.

\section{Discussion}
\label{sec:discussion}

We have presented a new algorithm for solving the Gaussian constrained
realization system for high-resolution CMB data. This method is based
on ideas from multi-grid (or multi-level) theory, and is fundamentally
different from the Conjugate Gradient methods traditionally used for
this problem. Being only weakly dependent on the signal-to-noise ratio
of the data set under consideration, our new method converges 
exponentially to numerical precision when properly tuned, and is capable 
of producing constrained realizations for the full resolution of a 
Planck-like data set within minutes. For comparison, we have yet to 
achieve robust full-sky convergence with CG methods for the same data 
set. Indeed, this particular issue was the single most important 
obstacle preventing a full-resolution analysis of the Planck 2013 data 
release with the Commander code.

The ultimate goal of this line of work is to perform an exact global
Bayesian analysis of the high-resolution, high-sensitivity 
observations now being produced by CMB experiments, including component
separation as described by \citet{eriksen:2008a}. For this to be
successful, multi-frequency and multi-component analysis must be added
to the algorithm. Other complications, such as the possible 
asymmetry of the CMB on large scales \citep[e.g.][]{planck:2013c}, will 
also need to be taken into account.
As such, the present paper represents only the first
step towards a complete solution. We also emphasize that the
algorithm as presented here is only the first implementation of a more
general framework, and we expect that many improvements with respect 
to computational speed, application to more general cases, overall 
robustness and stability, and even user interfaces, will be
introduced in the near future. Before concluding this paper, we will
mention a few relevant ideas, but leave all details for future
publications.

Firstly, as is evident from Figure \ref{fig:maskeffect}, our method is
quite sensitive to the behavior of the tails of the instrumental
beams extending as far out as the $10^{-5}$ level, as these
formally constrain the solution inside the mask. These tails are
not realistically known to such high accuracy, and so this issue is
therefore a modeling problem as well as a numerical problem. In
practice, it seems that in the absence of other options, one should 
just choose a form for the tails that falls quickly enough to not have 
an effect on the solution, and that allows a small computational 
bandlimit, $\lmax$. In
short, optimally tuning the tails of the beam profile may render a
more stable solution at a lower computational cost.

For an exact analysis of data from current and forthcoming
CMB experiments, one would ideally like to
account for the effect of asymmetric beams. With the above in mind, we
envision two solutions for this. One option is to modify the algorithm
so that the beams are defined in pixel space, as is done in FEBECop 
\citep{mitra:2011} for instance, and then carry the FEBECop beams
through to the computation of the smoother. The main challenge in this
scenario is how to avoid very expensive matrix-vector multiplications
at the coarse levels. Alternatively, and perhaps more simply, one could 
use the multi-level solver for perfect symmetric beams described
here as a preconditioner for a CG search, which then accounts for the
beam asymmetries in its own internal matrix multiplications.

Correlated noise is another significant complication for 
current CMB observations. While these
correlations have a complicated morphology in pixel space, being
convolved with the scanning strategy of the experiment, they are
simple to describe in the time-domain. With the vastly improved
convergence rate of the multi-level method presented here --- requiring 
only a handful of iterations to reach sub-$\mu$K errors --- it may for 
the first time be realistic to define the constrained realization system in
time-domain, rather than map-domain. As for asymmetric beams, this can
either be done by defining the multi-level scheme directly in
time-domain, or, if that does not succeed, by using the multi-level
solver for uncorrelated noise as a preconditioner for a time-domain CG
search. Going to time-domain also provides a direct route to handling
beam asymmetries and optical sidelobes by full-sky convolution
\citep{wandelt:2001}.

The current computational bottleneck in our implementation is the time 
needed to precompute the error smoothers. The time is spent almost 
exclusively on sampling rotationally-invariant operators at every position 
on the sphere by brute force evaluation of Equation
\eqref{eq:addition-theorem}. While the code for this computation is already 
highly optimized, as mentioned above, we do not exploit any
symmetries from pixel to pixel. The grid used within the multi-level
process is arbitrary, and not necessarily related to the grid of the
inverse-noise map, $\N$, or data vector, $\d$. A future implementation
of the algorithm will therefore employ a different grid with greater
symmetry than the HEALPix grid, which will only require evaluation of
the smoother blocks 3--7 times per pixel ring, thus reducing the
computational scaling from $O(k^2 \lmax \Npix)$ to $O(k^2 \lmax
\sqrt{\Npix})$.

Finally, the error smoother evaluation is currently not parallelized,
and only executes on a single CPU core. As the error smoothers only
need to work well for the local couplings, we expect to be able to
partition the sphere into multiple partially-overlapping domains, and
apply an error smoother on each domain in parallel, at the cost of
some extra computation on the domain borders. Assuming that this
approach is successful, the spherical harmonic transforms will
once again become the bottleneck of the overall algorithm.

\begin{acknowledgements}
  We thank Mikolaj Szydlarski and Martin Reinecke for useful
  discussions.  DSS, HKE and PB are supported by European Research
  Council grant StG2010-257080. KAM is supported by the Research
  Council of Norway through a Centre of Excellence grant to the Centre
  for Biomedical Computing at Simula Research Laboratory.

\end{acknowledgements}


\end{document}